%% file: root.tex
\documentclass[a4,10pt,twocolumn]{scrartcl}
\pdfoutput=1
\usepackage[top=2.4cm, bottom=2.4cm, left=1.5cm, right=1.5cm]{geometry}

\usepackage[hang]{footmisc}
\setfnsymbol{wiley}

\makeatletter
\renewcommand{\@biblabel}[1]{[#1]\hfill}
\makeatother

\usepackage{fancyhdr}
\fancypagestyle{firststyle}
{
  \fancyhf{}
  \fancyfoot[C]{This work has been submitted to Automatica for possible publication.}
}

\setkomafont{section}{\Large}
\setkomafont{subsection}{\large}
\setkomafont{subsubsection}{\normal}

\usepackage[format=plain,  
labelsep=period,           
font=small,                
labelfont=bf,              
skip=5pt                  
]{caption}
\usepackage{cite}
\usepackage{amsmath,amssymb,amsfonts}
\usepackage{algorithmic}
\usepackage{subcaption}
\usepackage{graphicx}
\usepackage{xfrac}
\usepackage{enumerate}
\usepackage{textcomp}
\usepackage{tikz} % <--- if tikz pictures are used
\usepackage{pgfplots}
\usepackage{xfrac}
\usepackage{mathptmx}
\usepackage{amsthm}
\DeclareMathAlphabet{\mathcal}{OMS}{cmsy}{m}{n}
\usepackage[colorlinks=true,%
linkcolor=black,%
urlcolor=black,%
citecolor=black]{hyperref}
\usepackage{bbm}
\usetikzlibrary{positioning}
\usetikzlibrary{calc}
\usetikzlibrary{decorations.markings,patterns,decorations.pathmorphing}
\usetikzlibrary{arrows, shapes}
\tikzstyle{block} = [draw, thick, node distance=0.5cm, minimum width=1cm, inner sep=6pt]
\tikzstyle{sum} = [draw, thick, circle, node distance=1cm, inner sep=3.5pt, path picture={\node at (path picture bounding box.center) [draw, anchor = center] {$+$};}]

\newtheorem{thm}{Theorem}
\newtheorem{assum}[thm]{Assumption}
\newtheorem{cor}[thm]{Corollary}
\newtheorem{defn}[thm]{Definition}
\newtheorem{exmp}[thm]{Example}

\newtheorem{lem}[thm]{Lemma}
\newtheorem{rem}[thm]{Remark}
\newenvironment{pf}{\paragraph{Proof:}}{}

\newcommand{\citep}[2][x]{\ifx x#1 \cite{#2}\else \cite[#1]{#2}\fi}
\newcommand{\R}{\mathbb{R}}
\newcommand{\N}{\mathbb{N}}
\newcommand{\I}{\mathbb{I}}
\newcommand{\X}{\mathbb{X}}
\newcommand{\U}{\mathbb{U}}
\newcommand{\Kinf}{\mathcal{K}_\infty}
\newcommand{\KL}{\mathcal{KL}}
\newcommand{\Pistar}{{\Pi^{\star}}}
\newcommand{\PiXstar}{{\Pi^{\star}_{\X}}}
\newcommand{\PiUstar}{{\Pi^{\star}_{\U}}}
\let\itshapeold\itshape
\renewcommand{\emph}[1]{{\itshapeold #1}}
\usepackage{tikz}
\usetikzlibrary{positioning}
\usetikzlibrary{calc}
\usetikzlibrary{decorations.markings,calc,patterns,decorations.pathmorphing}
\usetikzlibrary{arrows, shapes}
\newcommand{\refeq}[2]{\overset{\makebox[0pt][c]{\scriptsize #1}}{#2}}   
\newcommand{\new}[1]{#1}

\usepackage{authblk}

\usepackage{ulem}

\newcommand{\titlename}{Linearly discounted economic MPC without terminal conditions for periodic optimal operation}

\begin{document}\normalem
\title{\titlename
  \thanks{
    F. Allgöwer and M. A. Müller are thankful that this work was funded by Deutsche Forschungsgemeinschaft (DFG, German Research Foundation) – AL 316/12-2 and MU 3929/1-2 - 279734922.  L. Schwenkel thanks the International Max Planck Research School for Intelligent Systems (IMPRS-IS) for supporting him.\newline
    $^1$ L. Schwenkel and F. Allgöwer are with the Institute for Systems Theory and Automatic Control, University of Stuttgart, Stuttgart 70550, Germany (e-mail: $\{$schwenkel, allgower$\}$@ist.uni-stuttgart.de).\newline
    $^2$ A. Hadorn is with ISG Industrielle Steuerungstechnik GmbH, 70563 Stuttgart, Germany. (email: alexanderhadorn@mail.de).\newline   
    $^3$ M. A. Müller is with the Institute of Automatic Control, Leibniz University Hannover, 30167 Hannover, Germany (email: mueller@irt.uni-hannover.de).}}
\author{Lukas Schwenkel$^\text{1}$, Alexander Hadorn$^\text{2}$, Matthias A. Müller$^\text{3}$, and Frank Allgöwer$^\text{1}$}

\date{\vspace{-1cm}}

\maketitle
\thispagestyle{firststyle}

\begin{abstract}
\textbf{Abstract: }\input{abstract.tex}

\end{abstract}

\input{intro.tex}

\input{setup.tex}
\input{mpc.tex}
\input{results.tex}
\input{stability.tex}
\input{example.tex}
\input{conclusion.tex}
\input{main_proofs.tex}

{\small
\bibliographystyle{plain}
\bibliography{../../../IST_Literatur/my_bib}}

\end{document}

%% file: abstract.tex
In this work, we study economic model predictive control (MPC) in situations where the optimal operating behavior is periodic.
In such a setting, the performance of a \new{standard} economic MPC scheme without terminal conditions can generally be far from optimal even with arbitrarily long prediction horizons.
Whereas there are modified economic MPC schemes that guarantee optimal performance, all of them are based on prior knowledge of the optimal period length or of the optimal periodic orbit itself.
In contrast to these approaches, we propose to achieve optimality by multiplying the stage cost by a linear discount factor.
This modification is not only easy to implement but also independent of any system- or cost-specific properties, making the scheme robust against online changes therein.
Under standard dissipativity and controllability assumptions, we can prove that the resulting linearly discounted economic MPC without terminal conditions achieves optimal asymptotic average performance up to an error that vanishes with growing prediction horizons.
Moreover, we can guarantee practical asymptotic stability of the optimal periodic orbit \new{under the additional technical assumption that dissipativity holds with a continuous storage function}.
We complement these qualitative guarantees with a quantitative analysis of the transient and asymptotic average performance of the linearly discounted MPC scheme in a numerical simulation study.

%% file: intro.tex
\section{Introduction}
Economic model predictive control (MPC) (see, e.g., \citep{Ellis2014}, \citep{Gruene2017}, \citep{Faulwasser2018}) is an appealing control strategy for process control and other engineering applications due to its ability to directly optimize an economic criterion while ensuring constraint satisfaction.
In economic MPC, the control input is computed at each time step by solving a finite-horizon optimal control problem (OCP), in which the cost to be optimized can represent energy consumption, production amounts, or other physical or virtual costs.
In such a setting with a general stage cost, the optimal operating behavior is not explicitly specified but only implicitly through the interplay of the system, the constraints, and the stage cost resulting in a steady state, some periodic orbit, or an even more general trajectory (compare \cite{Angeli2012}). 
The optimization-based control strategy of economic MPC seems intuitive and is also very successful in some examples, however, in general, its closed-loop performance can be far from optimal.
Hence, there is a need for modifications of the OCP to obtain mathematical closed-loop performance guarantees.

The most common modification in MPC is the use of terminal conditions (i.e., terminal costs and/or terminal constraints).
If the optimal operating behavior is known to be a steady state, then \new{its} asymptotic stability can be guaranteed with suitable terminal conditions under certain dissipativity and controllability assumptions (see \cite{Amrit2011}).
The same holds true if the optimal operating behavior is a periodic orbit and the terminal conditions are suitably adapted (see \cite{Zanon2017}).
Whereas the use of terminal conditions leads to an optimal asymptotic average performance of the closed loop, it requires significant offline design efforts including knowledge of the optimal operating behavior and of a local control Lyapunov function with respect to it.
Furthermore, as soon as the cost changes online a redesign of the terminal conditions is needed to maintain the performance and stability guarantees.
Therefore, it is much more practicable to implement the OCP without terminal conditions.
Under similar assumptions and if the optimal operating behavior is a steady state, then such \new{an MPC scheme without terminal conditions} achieves suboptimal asymptotic average performance, where the suboptimality is vanishing with growing prediction horizons (see \citep{Gruene2013}, \citep{Gruene2014}).
However, this result does not hold in general if the optimal operating behavior is periodic as \cite[Ex.~4]{Mueller2016} observed: 
If the value function varies on the optimal periodic orbit, an unrewarding first step (e.g., waiting) may be preferred just to have a certain phase at the end of the prediction horizon. In closed-loop MPC, however, only this first step is implemented, which may result in a severe performance loss. 
This problem can even occur with arbitrarily large prediction horizons.
As \new{a} solution, \cite{Mueller2016} propose to implement a $p^\star$-step MPC scheme, where $p^\star$ is the optimal period length.
Alternatively, \cite{Koehler2018} require the stage cost and the value function to be constant on the optimal periodic orbit.
However, both solutions are not entirely satisfying since they either only work in a particular special case or still depend on the system\new{-} and cost\new{-}specific knowledge of the optimal period length such that an offline design is necessary and needs to be repeated whenever the system or the economic cost change during operation.
Moreover, the  $p^\star$-step MPC scheme leads only to convergence but not to stability guarantees.
Therefore, we propose in this work a novel approach that does not suffer from these drawbacks.

In particular, we propose to use a linearly discounted economic MPC scheme without terminal conditions (LDE-MPC).
With this discount, we mitigate the troubling effects at the end of the prediction horizon, whereas we do not require any offline design since this discount factor is independent of any system\new{-} or cost\new{-}specific property.
The main contribution of this work is to prove optimal asymptotic average performance up to an error vanishing with growing prediction horizons as well as practical asymptotic stability.
We establish these results based on a weaker version of the well\new{-}known turnpike property, which is commonly used to analyze economic MPC without terminal conditions (see, e.g., \cite{Gruene2017}).
Further, we complement the qualitative guarantees with a quantitative analysis of the transient and asymptotic average performance of the linearly discounted MPC scheme in a numerical simulation study and compare it to the undiscounted $1$-step and $p^\star$-step MPC scheme.

We want to emphasize that the goal of the proposed linearly discounted MPC is to solve the \emph{undiscounted} infinite-horizon optimal control problem. 
This is in contrast to related works on discounted economic MPC (e.g., \citep{Wuerth2013}, \citep{Gruene2016}, \citep{Gruene2021a}, \citep{Gruene2021}, and \citep{Zanon2022}), where exponential discounts are used in the MPC scheme to solve the \emph{exponentially discounted} infinite-horizon optimal control problem.
In our work, we show that \emph{linear} discounts enable economic MPC without terminal conditions to solve the undiscounted infinite-horizon optimal control problem not only when the optimal operating behavior is a steady state but also when it is a periodic orbit.
Although linear discounts are much less common than exponential discounts, they have been used before in MPC by \cite{Soloperto2022}, however, in a different context of learning-based MPC.

This article is structured as follows: After denoting the problem setup more formally in Sec.~\ref{sec:setup} and defining the discounted OCP in Sec.~\ref{sec:mpc}, we \new{show recursive feasibility in Sec.~\ref{sec:feasibility} and} state the performance result in Sec.~\ref{sec:results} and the stability result in Section~\ref{sec:stability}.
To improve readability, we moved the technical parts of the proofs of these results to the last Section~\ref{sec:proofs} after the numerical analysis in Section~\ref{sec:exmp} and conclusion in Section~\ref{sec:conclusion}.

\textbf{Notation.}
We denote the set of naturals including $0$ by $\N$, the set of reals by $\R$, and the set of integers in the interval $[a,b]$ by $\I_{[a,b]}$ for $a\leq b$ and define $\I_{[a,b]}=\emptyset$ for $a>b$.
Further, we define the notation $[k]_p$ for the modulo operation, i.e., for the remainder when dividing $k$ by $p$.
For $x\in\R$, the floor operator $\lfloor x \rfloor$ crops all decimal places.
Let $\alpha,\delta :[0,\infty)\to[0,\infty)$ be continuous functions. We say $\alpha \in \mathcal K_\infty$ if and only if $\alpha$ is monotonically increasing functions and satisfies $\alpha(0)=0$ and $\lim_{t\to\infty} \alpha(t) = \infty$.
Moreover, we say $\delta \in \mathcal L $ if and only if $\delta$ is monotonically decreasing and satisfies $\lim_{t\to\infty} \delta(t) = 0$.

%% file: setup.tex
\section{Problem setup}\label{sec:setup}
Consider the nonlinear discrete-time system
\begin{align}\label{eq:sys}
  x(k+1)=f(x(k),u(k))
\end{align}
subject to the constraints $x(k)\in\X\subset \R^n$ and $u(k)\in\U\subset \R^m$.
We denote the trajectory resulting from a specific input sequence $u \in \U^T$ \new{of length $T\in\N$} and the initial condition $x_0\in\X$ with $x_u(k,x_0)$, which is defined by $x_u(0,x_0)=x_0$ and $x_u(k+1,x_0)=f(x_u(k,x_0),u(k))$ for $k\in\I_{[0,T-1]}$.
Occasionally, we will use this notation also for feedback laws $\mu:\X \to \U$, in the natural sense $u(k)=\mu(x_\mu(k,x_0))$.
Further, for each $x\in\X$ we denote the set of feasible control sequences of length $T$ starting at $x$ with $\U^T(x):=\{u\in\U^T \mid \forall k\in \I_{[0,T]}: x_u(k,x)\in\X\}$.
The system is equipped with a stage cost function $\ell:\X\times \U\to \R$ and the control objective is to minimize the resulting asymptotic average performance 
\begin{align}
  J_\infty^\mathrm{av}(x,u) := \limsup_{T\to\infty} \frac 1 T J_T(x,u)
\end{align}
where for each $x\in\X$ and $u\in\U^T(x)$, the accumulated cost $J_T(x,u)$ is defined by
\begin{align}
  J_T(x,u):=\sum_{k=0}^{T-1} \ell (x_u(k,x),u(k)).
\end{align}
One of the main contributions of this article is to prove that we obtain optimal asymptotic average performance \new{up to an error that vanishes with growing prediction horizons}.
The assumptions we need to establish this result are listed in the remainder of this section. 
We want to emphasize that \new{we do not need any additional assumptions compared} to the ones used in \cite{Mueller2016} to prove a similar performance bound for an undiscounted $p^\star$-step MPC scheme.

\begin{assum}[Continuity and compactness]\label{ass:tech}\ \\
  The functions $f$ and $\ell$ are continuous, and the constraints $\X\times \U$ are compact. 
\end{assum}

Let us formally define (optimal, minimal) periodic orbits\footnote{We use $p$-tuples $\Pi\in (\X \times \U)^p$ as in \citep{Zanon2017} instead of subsets $\Pi\subseteq \X \times \U$ with $p$ elements as in \citep{Mueller2016}. The definition of \emph{minimal} orbits, however, is analogous to \citep{Mueller2016} to not only exclude multiple laps of the same orbit as in \citep{Zanon2017} but to also exclude, e.g., $8$-shaped orbits.}.

\begin{defn}[Optimal periodic orbit]\label{def:orbit}\ \\
  A $p$-tuple $\Pi \in (\X\times \U)^p$, $p\in\N$  is called a {\normalfont feasible $p$-periodic orbit}, if its projection $\Pi_\X$ onto $\X^p$ satisfies 
  \begin{align}
    \Pi_\X ([k+1]_p) = f(\Pi (k) )
  \end{align}
  for all $k \in \I_{[0,p-1]}$.
  A $p$-periodic orbit $\Pi$ is called minimal, if $\Pi_\X(k) = \Pi_\X(j) \Rightarrow k=j$ for all $k,j \in \I_{[0,p-1]}$.
  The distance of a pair $(x,u)\in\X\times \U$ to the orbit $\Pi$ is defined as $\|(x,u)\|_\Pi := \inf_{k\in \I_{[0,p-1]}} \|(x,u)-\Pi(k)\|$.
  The set of all feasible $p$-periodic orbits is denoted by $S_\Pi^p$.
  The average cost at $\Pi \in S_\Pi^p$ is defined as $\ell^p(\Pi):=\frac 1 p \sum_{k=0}^{p-1} \ell (\Pi(k))$.
  If a feasible $p^\star$-periodic orbit $\Pistar$ satisfies 
  \begin{align}\label{eq:ellstar}
    \ell^{p^\star} (\Pistar) = \inf_{p \in \N,\,\Pi\in S_\Pi^{p}} \ell^p( \Pi)=: \ell^\star,
  \end{align}
  then $\Pistar$ is called an {\normalfont optimal periodic orbit} and $p^\star$ is called an {\normalfont optimal period length}.
\end{defn}
Note that Ass.~\ref{ass:tech} guarantees that $\ell^\star$ in \eqref{eq:ellstar} is finite.
\new{Further, note that in general there may exist multiple or no optimal orbits $\Pistar$.}
However, if the following assumption of strict dissipativity (taken from \citep[Ass.~1]{Koehler2018}) is satisfied for a minimal orbit $\Pi^\star$, then this orbit is optimal and is the unique optimal orbit up to phase shifts. 
Further, strict dissipativity implies that the system is optimally operated at a periodic orbit, i.e., the best achievable asymptotic average performance is $\ell^\star$ \citep{Mueller2015a}.
\begin{assum}[Strict dissipativity]\label{ass:diss}\ \\
  There exist a storage function $\lambda :\X \to \R$\new{, a bound $\bar \lambda\in\R$ with $|\lambda(x)|\leq \bar \lambda$ for all $x \in \X$,} and a function $\underline\alpha_{\tilde \ell} \in \Kinf$, such that the rotated stage cost
  \begin{align}\label{eq:rot_cost}
    \tilde\ell(x,u)=\ell(x,u)-\ell^\star+\lambda(x)-\lambda(f(x,u))
  \end{align}
  satisfies for all $x\in\X$ and all $u\in\U^1(x)$
  \begin{align}\label{eq:strict_diss}
    \tilde\ell(x,u)\geq \underline \alpha_{\tilde\ell}(\|(x,u)\|_{\Pistar}).
  \end{align}
\end{assum}
Additionally, we need the following two controllability conditions taken from \citep[Ass.~10 and~11]{Mueller2016}.
\begin{assum}[Local controllability at $\Pistar$]\ \label{ass:loc_ctrb}\\
  There exists $\kappa >0, M'\in \N$ and $\rho \in \Kinf$ such that for all $z\in\PiXstar$ and all $x,y \in \X$ with $\|x-z\|\leq \kappa$ and $\|y-z\|\leq \kappa$ there exists a control sequence $u\in \U^{M'}(x)$ such that $x_u(M',x)=y$ and 
  \begin{align}\label{eq:loc_ctrb}
    \|(x_u(k,x),u(k))\|_{\Pistar} \leq \rho\big(\max \{\|x\|_\PiXstar,\|y\|_\PiXstar\}\big)
  \end{align}
  holds for all $k\in\I_{[0,M'-1]}$.
\end{assum}
\new{In the following, we will consider initial conditions $x_0\in\X_0 \subseteq \X$ and we assume that the optimal orbit $\PiXstar$ can be reached from all $x_0\in \X_0$ in finite time\footnote{Technically, Ass.~\ref{ass:reachability} guarantees finite-time reachability of a neighborhood of $\Pistar$. Together with the local controllability Ass.~\ref{ass:loc_ctrb}, reachability of $\Pi^\star$ in $M''+M'$ steps follows.}.}
\begin{assum}[Finite-time reachability %
  of $\Pistar$]\label{ass:reachability}\ \\
For $\kappa >0$ from Ass.~\ref{ass:loc_ctrb} there exists $M''\in\N$\new{, $M''\geq 1$,} such that for each \new{ $x_0\in \X_0$} there exists $K\in \I_{[0,M'']}$ and $u\in\U^K(x)$ such that $\|x_u(K,x)\|_\PiXstar \leq \kappa$.
\end{assum}
\new{Verifying Assumption~\ref{ass:diss} in practice is discussed in~\cite{Berberich2020}. A sufficient condition to verify Assumption~\ref{ass:loc_ctrb} is controllability of the linearization of the $p^\star$-step system $x_{t+1} = x_u(p^\star,x_t)$ at each point on the optimal orbit $\Pi_{\mathbb{X}}^\star(i)$, $i=\mathbb I_{[0,p^\star-1]}$, compare~\cite[Theorem~7]{Sontag1998a}.
Furthermore, checking Assumption~\ref{ass:reachability} for $x_0\in\X_0$ is an OCP with horizon length $M''$ and a terminal constraint. }

As shown in \cite[Cor.~2]{Koehler2018}, local controllability (Ass.~\ref{ass:loc_ctrb}) guarantees equivalence of the strict dissipativity assumption (Ass.~\ref{ass:diss}) and the one from \citep[Ass.~9]{Mueller2016}.
Hence, our setup is equivalent to \citep{Mueller2016} except that we do not use the simplifying assumption of control invariance of $\X$, but derive recursive feasibility in Sec.~\ref{sec:feasibility}.

%% file: mpc.tex
\section{Linearly discounted MPC scheme}\label{sec:mpc}

In this section, we define the LDE-MPC scheme starting with the finite-horizon discounted cost functional
\begin{align}\label{eq:costfct}
  J_T^{\beta_N} (x,u) := \sum_{k=0}^{T-1} \beta_N(k) \ell (x_u(k,x),u(k))
\end{align}
with the linear discount function
\begin{align}\label{eq:beta}
  \beta_N(k) := \frac{N-k}{N}
\end{align}
and $N\geq T$.
Almost always we consider $T=N$, therefore, we define $J_N^\beta:=J_N^{\beta_N}$ for ease of notation.
Further, the corresponding optimal value function is
\begin{align}\label{eq:valuefct}
  V^{\beta}_N (x) := \inf_{u\in\U^N(x)} J_N^{\beta} (x,u).
\end{align}
Due to Ass.~\ref{ass:tech} we know that $J_N^\beta$ is continuous and due to Ass.~\ref{ass:reachability} that for each $x\in\X_0$ the set $\U^N(x)$ is nonempty and compact for all $N\in\N$.
Therefore, there exists for each $x\in\X_0$ a possibly non-unique input sequence $u^\beta_{N,x}\in\U^N(x)$ that attains the infimum, i.e., $V_N^\beta (x) = J_N^\beta (x,u_{N,x}^\beta)$.
Then we can define the standard MPC feedback law
\begin{align}\label{eq:mpc_cl}
  \mu_N^\beta (x) := u_{N,x}^{\beta}(0),
\end{align}
that is, for a given $x$ we minimize $J_N^\beta (x,u)$ and take the first element of the minimizing sequence $u_{N,x}^\beta$ (if non-unique of any).
Note that in consistency with this notation we denote by $V_N^1$, $J_N^1$, $u_{N,x}^1$, and $\mu_N^1$ the well-known undiscounted (i.e., $\beta_N(k)=1$) value function, cost function, optimal input sequence, and resulting MPC feedback.
Moreover, we define the rotated cost functional $\tilde J_N^\beta$ and the rotated value function $\tilde V_N^\beta$ in the exact same way as $J_N^\beta$ and $V_N^\beta$, with the only difference that we take the rotated stage cost $\tilde \ell$ instead of $\ell$.
These two functions will play a crucial role in the performance and stability analysis of the MPC scheme.

Before we start analyzing the closed-loop performance of this scheme, we want to share some intuition how discounting can be beneficial when dealing with a periodic optimal behavior.
In~\cite[Example~4]{Mueller2016} it was observed that the optimal open-loop trajectories prefer to end at a specific phase on the optimal orbit such that any odd prediction horizon starts to wait one step, which leads in an MPC scheme to waiting forever as we recursively apply only the first step.
When using a discount factor these effects at the end of the prediction horizon become less important compared to the first time step and we can overcome this problem as the following example shows.

\begin{figure}
  \centering
  \resizebox{0.8\linewidth}{!}{\begin{tikzpicture}[inner sep=2pt, minimum width=0.8cm, node distance = 2cm]
    \node[circle,draw, fill=black!10] (a) {$1$};
    \node[circle,draw, fill=black!10, left =of a] (b) {$0$};
    \node[circle,draw, fill=black!10, left =of b] (c) {$-1$};
    \draw[-latex] (c) -- node[above]{\small$\ell(x,u)=1$} node[below] {\small$u=0$}  (b);
    \draw[-latex] (b) to [bend left=35]  node[above]{\small$\ell(x,u)=0$} node[below] {\small$u=1$}  (a);
    \draw[-latex] (a) to [bend left=35]  node[above] {\small $u=0$} node[below] {\small$\ell(x,u)=1.5$} (b);
    \draw[-latex] (c.north) arc[start angle=15, end angle=250, x radius=0.328cm, y radius=0.328cm]  node[above, yshift=0.2cm] {\small \begin{tabular}{c} $\ell(x,u)=1$  \phantom{xxx$+\varepsilon$}\\$u=-1$ \hspace{1.5cm}\  \end{tabular}} -- (c.west);
  \end{tikzpicture}}
\caption{Illustration of the states $x$ (nodes) and feasible transitions (edges) with
  corresponding input $u$ and cost $\ell$ in Example~\ref{exmp:motivation_disc}. This diagram is taken from \cite{Mueller2016}.}
\end{figure}
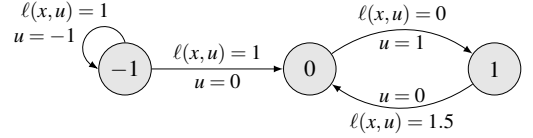
\begin{figure}
  \centering
  \resizebox{\linewidth}{!}{\input{fig_motiv_exmp.tex}}
  \caption{Comparison of the cost of strategies $u^1$ and $u^2$ for different $N$ depicted for both, the linearly discounted and the undiscounted cost functionals. A negative value indicates that strategy $u^1$ results in a better cost than strategy $u^2$.}
  \label{fig:motiv_exmp}
\end{figure}
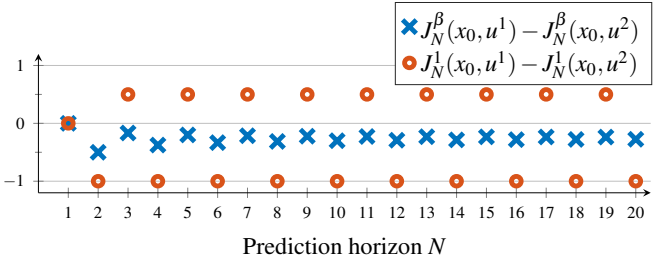

\begin{exmp}[Motivating example% -- linearly discounted
  ] \label{exmp:motivation_disc}\ \\
\emph{Consider the one-dimensional system $x(k+1) = u(k)$ depicted Fig.~\ref{fig:motiv_exmp}
  with state and input constraint set $\mathbb Z = \{(-1,-1),$ $(-1, 0),$ $(0, 1),$
  $(1, 0)\}$ consisting of four elements only and cost $\ell(x, u)$ defined as $\ell(-1,-1)=1$, $\ell(-1,0)=1$, $\ell(0,1) = 0$, $\ell(1,0) =1.5$, i.e., consider~\cite[Example~4]{Mueller2016} with $\varepsilon=0.5$. The system is optimally operated at the two-periodic orbit given by $\Pi^\star = ( (0, 1), (1, 0) ) $, and with average cost $\ell^\star =\frac 1 2 \sum_{k=0}^{1} \ell (\Pi^\star (k))=\frac{3}{4}$. 
  We use the linearly discounted cost functional from \eqref{eq:costfct}.
  To find the minimizing $u_{N,x_0}^\beta$ for $x_0=-1$ for all horizon lengths $N\in\N$, we compare the two candidate strategies: $u^1\in \U^\infty (x_0)$ going immediately to the optimal orbit; and $u^2\in\U^\infty(x_0)$ waiting one step before going to the optimal orbit. 
  As we can see in Fig.~\ref{fig:motiv_exmp}, strategy $u^1$ is optimal in the discounted cost for all $N\in\N$ and thus, the closed-loop input is $\mu_N^\beta (x_0)=u_{N,x_0}^\beta(0)=0$.
  Therefore, the resulting closed-loop trajectory converges for any prediction horizon $N\in\N$ to the optimal orbit in one step $x_{u_{N,x_0}^1}(1, x_0)=0$.
  With the undiscounted cost, we see that only for even prediction horizons $N\in 2\N$ the strategy $u^1$ is preferred, i.e., only for these $N$ there is closed-loop convergence to the optimal orbit (compare \cite[Example~4]{Mueller2016}).
}
\end{exmp}

%\new{TODO: paragraph moved to intro}
%The idea of discounting the stage cost function is not new, and even in the context of economic MPC, there are a few works considering exponential discounts, such as, for example, \citep{Gruene2016}, \citep{Gruene2021a}, \citep{Gruene2021}, \citep{Zanon2022}.
%However, the idea to use discounts in order to make economic MPC without terminal conditions applicable to larger classes of optimal operating behaviors is indeed novel; in fact all of these works only consider the case of optimal steady-state operation.
\new{It is worth to note that} an exponential discount $\beta_N(k)=\beta^k$ for some $\beta \in (0,1)$ would decrease too fast as we need $\lim_{N\to\infty} \sum_{k=0}^{N-1} \beta_N(k)=\infty$ to make sure that the reward in the discounted cost function of being at the optimal orbit is larger than any transient cost of approaching it as long as the prediction horizon $N$ is sufficiently large.
As the following sections show, a linear discount factor provides not only this property but is also appealing to analyze since we can exploit the linearity. 
%Linear discounts are much less common than exponential discounts, nonetheless, a similar linear discount factor has also been used by \cite{Soloperto2022}, however, in a different context of learning-based MPC.

%% file: fig_motiv_exmp.tex
\definecolor{mycolor1}{rgb}{0.00000,0.44700,0.74100}%
\definecolor{mycolor2}{rgb}{0.85000,0.32500,0.09800}%
\begin{tikzpicture}

\begin{axis}[%
width=3.5in,
height=0.8in,
at={(0.758in,0.481in)},
scale only axis,
axis x line=bottom,
axis y line=left,
xmin=0,
xmax=20.5,
ymin=-1.2,
ymax=1.2,
ytick={-1,0,1},
minor y tick num=1,
xtick={1,2,...,20},
xticklabel style = {font=\scriptsize},
yticklabel style = {font=\scriptsize},
xlabel=Prediction horizon $N$,
ymajorgrids,
legend style={legend cell align=left, align=left, draw=white!15!black, xshift = 0.2cm, yshift = 0.8cm}
]
\addplot [color=mycolor1, line width=2.0pt, only marks, mark=x, mark options={solid, mycolor1}, mark size = 4pt]
  table[row sep=crcr]{%
1	0\\
2	-0.5\\
3	-0.166666666666667\\
4	-0.375\\
5	-0.2\\
6	-0.333333333333333\\
7	-0.214285714285714\\
8	-0.3125\\
9	-0.222222222222222\\
10	-0.3\\
11	-0.227272727272727\\
12	-0.291666666666666\\
13	-0.230769230769233\\
14	-0.285714285714286\\
15	-0.233333333333334\\
16	-0.28125\\
17	-0.235294117647058\\
18	-0.277777777777778\\
19	-0.236842105263161\\
20	-0.274999999999999\\
21	-0.238095238095239\\
22	-0.272727272727273\\
23	-0.239130434782609\\
24	-0.270833333333334\\
25	-0.239999999999998\\
26	-0.269230769230772\\
27	-0.240740740740742\\
28	-0.267857142857144\\
29	-0.241379310344831\\
30	-0.266666666666667\\
31	-0.241935483870966\\
32	-0.265625\\
33	-0.242424242424244\\
34	-0.264705882352944\\
35	-0.242857142857146\\
36	-0.263888888888889\\
37	-0.243243243243244\\
38	-0.263157894736841\\
39	-0.243589743589743\\
40	-0.262500000000001\\
41	-0.243902439024385\\
42	-0.261904761904759\\
43	-0.244186046511622\\
44	-0.261363636363644\\
45	-0.244444444444433\\
46	-0.260869565217405\\
47	-0.244680851063837\\
48	-0.260416666666664\\
49	-0.244897959183671\\
50	-0.259999999999998\\
};
\addlegendentry{$J^\beta_N(x_0,u^1) - J^\beta_N(x_{0},u^2)$}

\addplot [color=mycolor2, line width=2.0pt, only marks, mark=o, mark options={solid, mycolor2}]
  table[row sep=crcr]{%
1	0\\
2	-1\\
3	0.5\\
4	-1\\
5	0.5\\
6	-1\\
7	0.5\\
8	-1\\
9	0.5\\
10	-1\\
11	0.5\\
12	-1\\
13	0.5\\
14	-1\\
15	0.5\\
16	-1\\
17	0.5\\
18	-1\\
19	0.5\\
20	-1\\
21	0.5\\
22	-1\\
23	0.5\\
24	-1\\
25	0.5\\
26	-1\\
27	0.5\\
28	-1\\
29	0.5\\
30	-1\\
31	0.5\\
32	-1\\
33	0.5\\
34	-1\\
35	0.5\\
36	-1\\
37	0.5\\
38	-1\\
39	0.5\\
40	-1\\
41	0.5\\
42	-1\\
43	0.5\\
44	-1\\
45	0.5\\
46	-1\\
47	0.5\\
48	-1\\
49	0.5\\
50	-1\\
};
\addlegendentry{$J^1_N(x_0,u^1) - J^1_N(x_{0},u^2)$}

\end{axis}

\end{tikzpicture}%

%% file: results.tex
\new{\section{Recursive feasibility}\label{sec:feasibility}
  In this section, we show that the linearly discounted economic MPC scheme without terminal conditions from Sec.~\ref{sec:mpc} is recursively feasible if initialized at $x_0\in \X_0$.
  It is known that recursive feasibility in economic MPC without terminal conditions can be established if optimal trajectories satisfy the so-called \emph{turnpike property} (see \cite{Faulwasser2015a}, \cite{Faulwasser2015b} for continuous time and \cite{Faulwasser2018} for discrete-time).
  The turnpike property} states that solutions of the OCP stay for all but a fixed number (independent of the length of the prediction horizon) of time steps in the neighborhood of the optimal behavior.
  Unfortunately, when discounting the stage cost we jeopardize this property as due to the small weights at the end of the horizon, more and more points could lie outside the neighborhood, hence, this number now depends on the length of the prediction horizon.
 \new{Still, we can show} that the number of points in the neighborhood grows faster than the number of points outside, which we therefore call the \emph{weak turnpike property}.
  
 \newcommand{\wtp}{Weak turnpike property}
 \newcommand{\rf}{Recursive feasibility}
  \new{
  \begin{defn}[\wtp]\label{defn:wtp}\ \\
    For $\varepsilon>0$, $N\in\N$ and $x\in\X$, define the number of points of the optimal trajectory $u_{N,x}^\beta$ in an $\varepsilon$-neighborhood of the optimal orbit $\Pi^\star$ as
    \begin{flalign} 
      &Q_\varepsilon^\beta (N,x) = \#\left\{k\in\I_{[0,N-1]}\ \Big| \left\|\big(x_{u_{N,x}^\beta}(k,x),u_{N,x}^\beta (k)\big)\right\|_{\Pistar} \leq \varepsilon \right\}.\!\!\!\!\!&
    \end{flalign}
    For $\alpha\in\Kinf$, $N_0\in\N$ we say that the OCP~\eqref{eq:valuefct} satisfies the weak turnpike property at $x\in\X$ if 
    \begin{align}\label{eq:wtp}
      Q_\varepsilon^\beta (N,x) \geq N - \frac{ \sqrt{N}}{\alpha(\varepsilon)}
    \end{align} 
    holds for all $N\geq N_0$ and all $\varepsilon>0$. We define the set of all points that satisfy the weak turnpike property as $\X_{\alpha,N_0}\subseteq \X$.
  \end{defn}}

Remember that in the commonly known turnpike property (see, e.g., \citep{Gruene2017}) the $\sqrt{N}$ term in \eqref{eq:wtp} is a constant independent of $N$. 
Hence, whereas the weak turnpike property does not imply that the number of points outside the $\varepsilon$-neighborhood $N-Q_\varepsilon^\beta (N,x)$ is bounded by a constant, it still satisfies that the proportion of points inside is growing to $1$, i.e., $\lim_{N\to\infty} \frac{1}{N} Q_\varepsilon^\beta (N,x)=1$.

\new{
The following theorem is the core of recursive feasibility, as it shows that there are positively invariant sets under the feedback $\mu_N^\beta$ for which the OCP~\eqref{eq:valuefct} is feasible, i.e., $V_N^\beta$ is finite. 
The crucial observation is that on this set the weak turnpike property holds and that the weak turnpike property can be used to construct a candidate solution for the next time point.
\begin{thm}[Weak turnpike property]\label{thm:wtp}
  Let Ass.~\ref{ass:tech},~\ref{ass:diss}, and~\ref{ass:loc_ctrb} hold and for $C\in\R$, $N_0\in\N$ let the set $\X_\mathrm{pi}(C,N_0)\subseteq \X$ be defined as the set of all $x\in\X$ that satisfy
  \begin{align}\label{eq:Xpi}
    V_N^\beta (x) - \frac{N+1}{2} \ell^\star +\lambda(x) + \bar \lambda \leq C
  \end{align}
  for all $N\geq N_0$. Then the weak turnpike property holds for all $x\in\X_\mathrm{pi}(C,N_0)$, i.e., $\X_\mathrm{pi}(C,N_0)\subseteq \X_{\alpha,N_0}$ with $\alpha(\varepsilon)=\frac{1}{\sqrt{2C}}\sqrt{\underline\alpha_{\tilde \ell}(\varepsilon)}$. Further, for each $C>C(M'):=M'(\ell_\mathrm{max}-\ell^\star)+2\bar \lambda +\ell^\star p^\star$ there exist $N_0\in\N$ such that $\X_\mathrm{pi}(C,N_0)$ is positively invariant under the MPC feedback law $\mu_N^\beta$ defined in \eqref{eq:mpc_cl} for all $N\geq N_0$.
\end{thm}
\begin{pf}
      To emphasize the structure of the proof, we outsource all technical steps to Lemma~\ref{lem:rcf},~\ref{lem:wtp_ub},~\ref{lem:wtp_lb}, and~\ref{lem:V_N_diff} in Section~\ref{sec:proofs}.
      The first crucial observation using Lemma~\ref{lem:rcf} and $|\lambda(x)|\leq \bar\lambda$ from Assumption~\ref{ass:diss} is that the left hand side of~\eqref{eq:Xpi} is an upper bound to $\tilde J_{N}^\beta (x,u_{N,x}^\beta)$ and hence, $\tilde J_{N}^\beta (x,u_{N,x}^\beta)\leq C$ for all $x \in \X_\mathrm{pi}(C,N_0)$ and all $N\geq N_0$.
      Such a bound on the rotated cost of optimal trajectories is sufficient for the weak turnpike property as shown in Lemma~\ref{lem:wtp_lb}, which yields $\X_\mathrm{pi}(C,N_0)\subseteq \X_{\alpha,N_0}$ with $\alpha (\varepsilon)=\frac{1}{\sqrt{2C}}\sqrt{\underline\alpha_{\tilde \ell}(\varepsilon)}$.
      Next we assume $C>C(M')$ and prove forward invariance of $\X_\mathrm{pi}(C,N_0)$ for some sufficiently large $N_0$, i.e., we show that $x_0\in\X_\mathrm{pi}(C,N_0)$ implies for all $N\geq N_0$ that $x_1=x_{\mu_N^{\beta}}(1,x_0)\in \X_\mathrm{pi}(C,N_0)$. 
      First, note that 
      \begin{align*}
        \tilde J_N^\beta(x_1,u_{N,x_1}^\beta)\leq V_N^\beta (x_1) - \frac{N+1}{2} \ell^\star +\lambda(x_1) + \bar \lambda
      \end{align*}
      as computed in Lemma~\ref{lem:rcf}.
      The second crucial observation is that the weak turnpike property $x_0\in \X_{\alpha,N_0}$ can be used to construct a feasible and almost optimal candidate solution for $x_{1}$ as shown in Lemma~\ref{lem:V_N_diff} which yields
      \begin{align*}
        &V_N^\beta (x_{1}) \leq V_N^\beta (x_0) +  \frac{N+1}{N} \left(\ell^\star - \ell(x_0,u_0) + \delta(N+1)\right)
      \end{align*}
      with $u_0=\mu_N^\beta(x_0)$ for all $N\geq N_0$ with $N_0$ from Lemma~\ref{lem:V_N_diff}.
      Further, using this inequality and Assumption~\ref{ass:diss} to upper bound $\ell^\star - \ell(x_0,u_0)$ yields
      \begin{align*}
        &V_N^\beta (x_{1}) - \frac{N+1}{2} \ell^\star +\lambda(x_{1}) + \bar \lambda \\
        &\leq V_N^\beta (x_0) - \frac{N+1}{2} \ell^\star +\lambda(x_0) - \frac{N+1}{N}\underline\alpha_{\tilde \ell}\left(\|(x_0,u_0)\|_\Pistar\right)\\ &\qquad + \bar \lambda + \frac{N+1}{N}\underbrace{\left(\frac{\lambda(x_0)-\lambda(x_{1})}{N+1}  + \delta(N+1)\right)}_{\leq \frac{2\bar \lambda}{N+1}+\delta(N+1) =: \sigma_1(N)}
      \end{align*}
      with $\sigma_1 \in \mathcal L$. 
      The third crucial observation is that by choosing long enough horizons $N\geq N_0$, the first step of the optimal trajectory goes either in the right direction, in the sense that the left hand side of~\eqref{eq:Xpi} decreases, or we are already so close to $\PiXstar$ that~\eqref{eq:Xpi} holds anyways.
      In particular, if $N_0$ from Lemma~\ref{lem:V_N_diff} is not already large enough, we increase it such that we have $\sigma_1(N_0) \leq \underline\alpha_{\tilde \ell}(\kappa)$ with $\kappa$ from Assumption~\ref{ass:loc_ctrb} and $\frac{N_0+1}{N_0} \sigma_1(N_0)\leq C-C(M')$, where the right hand side is positive since we assume $C>C(M')$.
      This choice enables us to treat the following two cases:
      \begin{itemize}
        \item case $\underline\alpha_{\tilde \ell}\left(\|(x_0,u_0)\|_\Pistar\right)\geq \sigma_1(N)$: Then due to $x_0\in \X_\mathrm{pi}(C,N_0)$ we have
        \begin{align*}
          &V_N^\beta (x_{1}) - \frac{N+1}{2} \ell^\star +\lambda(x_{1}) + \bar \lambda \\
          &\leq V_N^\beta (x_0) - \frac{N+1}{2} \ell^\star +\lambda(x_0) + \bar \lambda \\ 
          &\qquad + \frac{N+1}{N}\underbrace{\left(\sigma_1(N)-\underline\alpha_{\tilde \ell}\left(\|(x_0,u_0)\|_\Pistar\right)\right)}_{\leq 0} \leq C.
        \end{align*}
        \item case $\underline\alpha_{\tilde \ell}\left(\|(x_0,u_0)\|_\Pistar\right)< \sigma_1(N)$: Then $\|x_0\|_\PiXstar\leq \kappa$ as $\sigma_1(N) \leq \underline\alpha_{\tilde \ell}(\kappa)$. Thus, using Ass.~\ref{ass:loc_ctrb}, we can reach $\PiXstar$ in $M'$ steps. Using Lemma~\ref{lem:wtp_ub}, we obtain $V_N^\beta (x_0) - \frac{N+1}{2} \ell^\star +\lambda(x_{0}) + \bar \lambda\leq C(M')$.
        Hence, we have
        \begin{align*}
          &V_N^\beta (x_{1}) - \frac{N+1}{2} \ell^\star +\lambda(x_{1}) + \bar \lambda \\
          &\leq C(M') + \frac{N+1}{N}\big(\sigma_1(N)-\underline\alpha_{\tilde \ell}\left(\|(x_0,u_0)\|_\Pistar\right)\big) \leq C.
        \end{align*}
      \end{itemize}
      Hence, in both cases we have established $x_1\in\X_\mathrm{pi}(C,N_0)$ which proves that $\X_\mathrm{pi}(C,N_0)$ is indeed positively invariant. \hfill $\square$
\end{pf}
}

\new{
  Recursive feasibility for initial conditions $x_0\in\X_0$ is a direct corollary of Theorem~\ref{thm:wtp} as by compactness of $\X$ and $\U$ there is a $C\in \R$ such that~\eqref{eq:Xpi} holds for all $x\in\X_0$. 

  \begin{cor}[\rf] \ \label{cor:rf} \\
    Let Ass.~\ref{ass:tech},~\ref{ass:diss},~\ref{ass:loc_ctrb}, and~\ref{ass:reachability} hold. Then there exists $N_0\in\N$ such that $\X_0\subseteq \X_\mathrm{pi}(C_0,N_0)$ with $C_0=C(M''+M')$. In particular, the OCP~\eqref{eq:valuefct} is feasible for $x_{\mu_N^\beta}(k,x_0)$ and all $N\geq N_0$, $k\geq 0$.
  \end{cor}
  \begin{pf}
    As $M''\geq 1$ we have $C_0=C(M''+M')>C(M')$, hence $\X_\mathrm{pi}(C_0,N_0)$ is positively invariant by Theorem~\ref{thm:wtp}. Further, by Assumptions~\ref{ass:loc_ctrb} and~\ref{ass:reachability} we can reach $\PiXstar$ from any $x_0\in\X_0$ in at most $M'+M''$ steps. 
    Hence, by applying Lemma~\ref{lem:wtp_ub} we obtain $\X_0\subseteq \X_\mathrm{pi}(C_0,N_0)$. 
    Finally, OCP~\eqref{eq:valuefct} is feasible for all $x\in\X_\mathrm{pi}(C_0,N_0)$ as $V_N^\beta (x)$ is finite for such $x$ by~\eqref{eq:Xpi}. \hfill $\square$
  \end{pf}
}

\section{Optimal asymptotic average performance}\label{sec:results}

In this section, we show that the linearly discounted economic MPC scheme without terminal conditions from Sec.~\ref{sec:mpc} achieves an asymptotic average performance that is optimal up to an error vanishing with growing prediction horizons. 
This performance result is analogous to the results known from other economic MPC schemes without terminal conditions, compare \citep{Gruene2014} in case of optimal steady-state operation or \citep{Mueller2016} in case of optimal periodic operation using a $p$-step MCP scheme.
In these works, the proof of the performance bound is heavily based on \new{the turnpike property.
In the last section, we have seen that in our setup we can only guarantee the weak turnpike property.
Nonetheless, in the following result we see that the weak version is sufficient to prove an asymptotic average performance bound that is analogous to the one from \citep{Mueller2016}.}

\newcommand{\performance}{Asymptotic average performance}
\begin{thm}[\performance]\label{thm:performance}\ \\
  \emph{Let Ass.~\ref{ass:tech},~\ref{ass:diss}, and~\ref{ass:loc_ctrb} hold \new{and let $C>C(M')$}. Then there exists $\delta\in\mathcal L$, \new{$N_0\in\N$} such that for each prediction horizon length \new{$N\geq N_0$}, the MPC feedback law $\mu_N^\beta$ defined in \eqref{eq:mpc_cl} results in an asymptotic average performance that is \new{for all $x\in\X_\mathrm{pi}(C,N_0)$} not worse than
    \begin{align}\label{eq:performance}
      J_\infty^\mathrm{av} (x,\mu_N^\beta) \leq \ell^\star + \delta (N).
  \end{align}}
\end{thm}
\begin{pf}
  To clearly highlight the core idea, we outsource all technical steps to Lemma~\ref{lem:dpp} and~\ref{lem:V_N_diff} in Sec.~\ref{sec:proofs}.
  To begin with the proof, we use the dynamic programming \new{principle}~\eqref{eq:dpp} from Lemma~\ref{lem:dpp} to obtain 
  \begin{align*}
    &J_T(x,\mu_N^\beta ) = \sum_{k=0}^{T-1} \ell \left(x_{\mu_N^\beta}(k,x), \mu_N^\beta\Big(x_{\mu_N^\beta}(k,x)\Big)\right) \\
    &\refeq{\eqref{eq:dpp}}{=} \ \sum_{k=0}^{T-1}\left( V_N^\beta \Big(x_{\mu_N^\beta}(k,x)\Big) - \frac {N-1}N V_{N-1}^\beta \Big(x_{\mu_N^\beta}(k+1,x)\Big)\right)\\
    &= \ V_N^\beta (x) - \frac {N-1}N V_{N-1}^\beta \Big(x_{\mu_N^\beta}(T,x)\Big) \\&
    \quad \ + \sum_{k=1}^{T-1} \left(V_N^\beta \Big(x_{\mu_N^\beta}(k,x)\Big) - \frac {N-1}N V_{N-1}^\beta \Big(x_{\mu_N^\beta}(k,x)\Big)\right).
  \end{align*}
  The main part of this proof is to bound the difference of $V_N^\beta$ and $V_{N-1}^\beta$.
  \new{Due to Theorem~\ref{thm:wtp} and $x\in\X_\mathrm{pi}(C,N_0)$, the weak turnpike property $x_{\mu_N^\beta}(k,x)\in\X_{\alpha,N_0}$ holds for all $k\geq 0$. Hence, we can use it} to extend the solution of $V_{N-1}^\beta$ by one step to an almost optimal candidate solution for $V_N^\beta$, the technical details are proven in Lemma~\ref{lem:V_N_diff}.
  Applying this Lemma leads to
  \begin{align*}
    J_T(x,\mu_N^\beta )\ &\refeq{\eqref{eq:V_N_diff}}{\leq}\ V_N^\beta (x) - \frac {N-1}N V_{N-1}^\beta \Big(x_{\mu_N^\beta}(T,x)\Big) \\
    &\quad \ + (T-1) \big( \ell^\star + \delta(N)\big)\\
    &\leq V_N^\beta (x) + (T-1) \big( \ell^\star + \delta(N)\big)
  \end{align*}
  for all $N\geq N_0\new{+1}$, where the last inequality holds when we assume without loss of generality that $\ell$ and thus also  $V_{N-1}^\beta $ are non-negative.
  A justification for this assumption can be found in Rem.~\ref{rem:wlog_ell_nn}.
  Further, we compute the $\limsup$ as follows
  \begin{align*}
    J_\infty^\mathrm{av} (x,\mu_N^\beta) &= \limsup_{T\to\infty} \frac 1 T J_T(x,\mu_N^\beta)\\
    &\leq  \limsup_{T\to\infty} \frac 1 T \left( V_N^\beta (x) + (T-1) \big( \ell^\star + \delta(N)\big)\right) \\
    &= \ell^\star + \delta(N),
  \end{align*}
  \new{where we used that $V_N^\beta(x)<\infty$ for all $x\in\X_\mathrm{pi}(C,N_0)$ due to~\eqref{eq:Xpi}.}
  \hfill $\square$
\end{pf}

%% file: stability.tex
\section{Practical asymptotic stability}\label{sec:stability}

In this section, we prove practical asymptotic stability of the optimal orbit when applying the LDE-MPC scheme.

\begin{defn}[Practical asymptotic stability]\label{defn:stability}\ \\
  \new{Let $S\subseteq \X$ be a positively invariant set of system~\eqref{eq:sys} under the feedback $\mu:S\to\U$.} A $p$-periodic orbit $\Pi\in(\new{S}\times \U)^p$ of~\eqref{eq:sys} is called {\normalfont practically asymptotically stable} on $S$ w.r.t. $\varepsilon\geq 0$ under the feedback $\mu$ if there exists $\beta\in\KL$ such that for all $x\in S$ and $k\in \N$ it holds
  \begin{flalign}\label{eq:stability}
    \|(x_\mu (k,x), \mu(x_\mu (k,x)))\|_{\Pi} \leq \max \{\beta(\|x\|_{\Pi_\X},k),\varepsilon\}.\hspace{-1cm}\ &&
  \end{flalign}
\end{defn}

\begin{rem}\label{rem:stability}
  Definition~\ref{defn:stability} guarantees practical asymptotic stability in state and input, which implies for $\varepsilon$ small enough and $k$ large enough that the trajectory $x_\mu$ evolves along the sequence $\Pi_\X$ in the sense that for each $k$ large enough there exists $j\in\I_{[0,p-1]}$ such that $\|x_\mu(k,x)-\Pi_\X(j)\|\leq \varepsilon$ and $\|x_\mu(k+1,x)-\Pi_\X([j+1]_p)\|\leq \varepsilon$.
  This is different from stability of the set $\{\Pi_\X(j) \mid j\in \I_{[0,p-1]}\}$ and for non minimal orbits different from stability of the trajectory $\Pi_\X ([k]_p)$.
  See also \cite[Theorem 5.6]{Zanon2017}, where depending on the strictness of the dissipativity different stability formulations are obtained for economic MPC with terminal conditions.
\end{rem}

Furthermore, we need to extend the standard definition of practical Lyapunov functions for steady states (see, e.g., \cite[Definition 2.3]{Gruene2014}) to periodic orbits.

\begin{defn}[Practical Lyapunov function]\label{defn:lyap}\ \\
  \new{Let $S\subseteq \X$ be a positively invariant set of system~\eqref{eq:sys} under the feedback $\mu:S\to\U$ and}
  let $\Pi\in(\new{S}\times \U)^p$ be a $p$-periodic orbit of system \eqref{eq:sys}. 
  A function $V:\new{S}\to \R$ is a {\normalfont practical Lyapunov function} \new{on $S$} w.r.t. $\delta \geq 0$ for the orbit $\Pi$ and system \eqref{eq:sys} under the feedback $\mu$, if there exist $\underline\alpha_{V},\overline\alpha_{V},\alpha_{\Delta\! V}\in\Kinf$ such that 
  \begin{align}\label{eq:prac_lyap_pd}
    \underline\alpha_{V}(\|(x,\mu(x))\|_{\Pi})\leq V (x) \leq \overline\alpha_{V} (\|x\|_{\Pi_\X})
  \end{align}
  holds for all $x\in\X$ and
  \begin{align}\label{eq:prac_lyap_decr}
    V\left(f\big(x,\mu(x)\big)\right) - V (x) \leq - \alpha_{\Delta\! V}(\|(x,\mu(x))\|_\Pi) + \delta
  \end{align}
  holds for all $x\in \new{S}$.
\end{defn}

The reason for having $\|(x,\mu(x))\|_{\Pi}$ in the lower bound and $\|x\|_{\Pi_\X}$ in the upper bound of~\eqref{eq:prac_lyap_pd} is due to fact that in \eqref{eq:stability} we similarly have $\|(x_\mu(k,x),\mu(x_\mu(k,x)))\|_{\Pi}$ on the left hand side of the inequality and $\|x\|_{\Pi_\X}$ on the right hand side.
Analogous to the steady state case (see \cite[Definition 2.2, 2.3, Theorem 2.4]{Gruene2014}), a practical Lyapunov function from Def.~\ref{defn:lyap} implies practical asymptotic stability of periodic orbits (Def.~\ref{defn:stability}).

\newcommand{\lyapunov}{Lyapunov function $\Rightarrow$ stability}
\begin{thm}[\lyapunov]\label{thm:lyapunov}\ \\
  \new{Let $S\subseteq \X$ be a positively invariant set of system~\eqref{eq:sys} under the feedback $\mu:S\to\U$,} let $\Pi\in(S\times \U)^p$ be a $p$-periodic orbit of system \eqref{eq:sys} and let $V:\new{S}\to \R$ be a practical Lyapunov function \new{on $S$} w.r.t. $\delta \geq 0$ for the orbit $\Pi$ and system \eqref{eq:sys} under the feedback $\mu$. 
  Then $\Pi$ is practically asymptotically stable \new{on $S$} w.r.t. $\varepsilon = \underline\alpha_{V}^{-1}(\overline\alpha_{V} (\alpha_{\Delta\! V}^{-1}(\delta))+\delta)$ under the feedback $\mu$.
\end{thm}
\begin{pf}
  Note that $ \overline\alpha_{V} (\|x\|_{\Pi_\X}) \leq  \overline\alpha_{V} (\|(x,\mu(x)\|_{\Pi})$. 
  By plugging this into the upper bound of \eqref{eq:prac_lyap_pd}, we have a similar practical asymptotic stability setup as \cite{Gruene2014} or \cite{Gruene2017} and we can follow the proof of \cite[Theorem 2.4]{Gruene2014} step by step with the obvious modifications to obtain 
  \begin{align*}
    \|(x_\mu (k,x), \mu(x_\mu (k,x)))\|_{\Pi} \leq \max \{\tilde\beta(\|(x,\mu(x))\|_{\Pi},k),\varepsilon\}
  \end{align*}
  with $\varepsilon = \underline\alpha_{V}^{-1}(\overline\alpha_{V} (\alpha_{\Delta\! V}^{-1}(\delta))+\delta)$ and $\tilde \beta$ as constructed in the proof of \cite[Theorem 2.19]{Gruene2017}.
  Now, we define $\beta(r,t):= \tilde \beta(\underline\alpha_{V}^{-1}(\overline\alpha_{V}(r)),t)$ and observe
  \begin{align*}
    \tilde\beta(\|(x,\mu(x))\|_{\Pi},k) \refeq{\eqref{eq:prac_lyap_pd}}{\leq} \tilde\beta(\underline\alpha_{V}^{-1}(\overline\alpha_{V}(\|x\|_{\Pi_\X})),k) = \beta(\|x\|_{\Pi_\X},k),
  \end{align*}
  which gives the desired inequality \eqref{eq:stability}.\hfill $\square$
\end{pf}

As we will see, the rotated value function $\tilde V_N^\beta$ as defined below~\eqref{eq:mpc_cl} is a practical Lyapunov function \new{on $\X_\mathrm{pi}(C,N_0)$}. 
We only need one additional technical assumption to ensure continuity of $\tilde V_N^\beta$.
\begin{assum}[Continuous storage function]\label{ass:lambda_cont}\ \\
  Assume that the storage function $\lambda$ of Ass.~\ref{ass:diss} is continuous and that there exists $\overline \alpha_\lambda\in\Kinf$ such that for all $x\in \X$ and all $k \in \I_{[0,p^\star -1]}$ it holds
  \begin{align}
    |\lambda(x) - \lambda(\PiXstar(k))| \leq \overline \alpha_\lambda (\|x-\PiXstar(k)\|).
  \end{align}
\end{assum}
Now we are prepared to show that any arbitrarily small neighborhood of the optimal orbit can be stabilized if the prediction horizon is sufficiently large.

\newcommand{\stability}{Practical asymptotic stability}
\begin{thm}[\stability]\label{thm:stability}\ \\
  Let Ass.~\ref{ass:tech},~\ref{ass:diss},~\ref{ass:loc_ctrb}, and~\ref{ass:lambda_cont} hold, \new{ let $C>C(M')$,} and assume that $\Pistar$ is minimal. Then there exists $\varepsilon\in\mathcal L$ \new{and $N_0\in\N$} such that the optimal periodic orbit $\Pi^\star$ is practically asymptotically stable \new{on $\X_\mathrm{pi}(C,N_0)$} w.r.t. $\varepsilon(N)$ under the MPC feedback $\mu_N^\beta$ \new{for all $N\geq N_0$}.
  In particular, the rotated value function $\tilde V_N^\beta (x)$ is a practical Lyapunov function \new{on $\X_\mathrm{pi}(C,N_0)$}.
\end{thm}
\emph{Proof.}
\new{Due to Theorem~\ref{thm:wtp} we know that we can choose $N_0$ such that $\X_\mathrm{pi}(C,N_0)$ is positively invariant. Hence, te desired result is proven according to Theorem~\ref{thm:lyapunov} if} there exists $\delta_1 \in \mathcal L$ such that $\tilde V_N^\beta$ is a practical Lyapunov function \new{on $\X_\mathrm{pi}(C,N_0)$} w.r.t. $\delta_1(N)$ for the orbit $\Pistar$.
First, Lemma~\ref{lem:rot_val_ub} verifies that $\tilde V_N^\beta$ satisfies~\eqref{eq:prac_lyap_pd}.
Second, the decrease condition \eqref{eq:prac_lyap_decr} follows due to \eqref{eq:strict_diss} and \eqref{eq:mpc_cl}, if we manage to show
\begin{align}\label{eq:prac_lyap_decr2}
  \tilde V_N^\beta\left(f\big(x,\mu_N^\beta (x)\big)\right)\! - \tilde V_N^\beta (x)  \leq -\tilde \ell \big(x, \mu_{N}^\beta(x)\big)  + \delta_1(N) \!\!\!\!\!\!\!&&
\end{align}
for some $\delta_1 \in \mathcal L$.
We conveniently denote $y=f\big(x,\mu_N^\beta (x)\big)$\new{, remark that $y\in\X_\mathrm{pi}(C,N_0)$ due to the positive invariance of $\X_\mathrm{pi}(C,N_0)$,} and rewrite \eqref{eq:prac_lyap_decr2} as
\begin{align}
  \tilde V_N^\beta(y)&-\lambda(y) - \tilde V_N^\beta (x)+\lambda(x) \leq -\ell \big(x, \mu_{N}^\beta(x)\big) +\ell^\star  + \delta_1(N). \label{eq:prac_lyap_decr3}
\end{align}
In order to prove this inequality, we exploit the weak turnpike property, which allows us to relate the left hand side of \eqref{eq:prac_lyap_decr3} to $V_N^\beta(y) - V_N^\beta (x)$.
Lemma~\ref{lem:sim_cost} carries out the rigorous derivation of this relation and yields
\begin{align}\label{eq:sim_cost1}
  \tilde V_N^\beta(y)-\lambda(y) - \tilde V_N^\beta (x)+\lambda(x)  \refeq{\eqref{eq:sim_tail_cost}}{\leq}  V_N^{\beta}(y)-  V_N^{\beta}(x) + \delta_2(N).
\end{align}
Recall $y=f\big(x,\mu_N^\beta (x)\big)$ and use Lemma~\ref{lem:V_N_diff} and~\ref{lem:dpp} to see
\begin{align*}
  -  V_N^{\beta}(x) &\refeq{\eqref{eq:dpp}}{=} - \ell\big(x,\mu_N^\beta(x)\big)-\frac{N-1}{N} V_{N-1}^\beta (y) \\
  &\refeq{\eqref{eq:V_N_diff}}{\leq} -V_N^\beta (y)- \ell\big(x,\mu_N^\beta(x)\big)+\ell^\star + \delta(N).
\end{align*}
Plugging this inequality into \eqref{eq:sim_cost1} and defining $\delta_1:= \delta+\delta_2\in\mathcal L$ yields \eqref{eq:prac_lyap_decr3}.
\hfill $\square$

\begin{rem}
  The additional assumption of minimality of $\Pistar$ is often taken for simplicity, e.g., in \cite{Mueller2016} and \cite{Koehler2018}. 
  Compared to~\cite{Mueller2016}, we were able to remove the assumption on minimality for the performance result in Theorem~\ref{thm:performance}.
  Removing it for the stability result in Theorem~\ref{thm:stability} poses some technical challenges, especially in the proof of Lemma~\ref{lem:sim_cost}.
  In particular, when estimating the sum $\Sigma_6$, the terms with $\lambda$ do not cancel out anymore.
  Nonetheless, we conjecture that it may be possible to prove practical asymptotic stability also for non-minimal orbits.
\end{rem}

%% file: example.tex
\section{Numerical Analysis}\label{sec:exmp}
In this section, we present two simulation examples of the proposed LDE-MPC scheme, in order to ...
\begin{enumerate}[(i)]
  \item show that \new{an undiscounted $1$-step} MPC scheme without terminal conditions can fail in more common setups with a connected constraint set (Example~4 and~18 in \citep{Mueller2016} are rather artificial).
  \item compare the asymptotic average performance of the discounted and undiscounted schemes.
  \item compare the transient performance of the discounted and undiscounted schemes.
\end{enumerate}
Whereas the first example achieves all of these goals and shows the benefits of LDE-MPC, the second example shows a limitation of the approach.

\begin{exmp}[Harmonic oscillator]\label{exmp:oscillator}\ \\
Consider a discrete-time two dimensional harmonic oscillator $x(k+1)=\left[\begin{smallmatrix} c & -s \\ s & c\end{smallmatrix}\right]x(k)+\frac 1 {\omega_0} \left[\begin{smallmatrix} s & c-1 \\ 1-c & s \end{smallmatrix}\right] u(k)$ with $c=\cos(h\omega_0)$, $s=\sin(h\omega_0)$, state $x(t) \in \R^2$, input $u(t)\in \R^2$, eigenfrequency $\omega_0 = 2\pi/6$, and discretization\footnote{The discrete-time dynamics can be obtained via exact discretization from the continuous-time oscillator $\dot x=Ax+Bu$ with $A=\left[\begin{smallmatrix} 0 & - \omega_0 \\ \omega_0 & 0\end{smallmatrix}\right]$ and $B=I$.} step size $h=1$.
The box constraints $\|x\|_\infty \leq x_\mathrm{max}=1$ and $\|u\|_\infty \leq u_\mathrm{max}= 0.1$  must be satisfied at all times, while minimizing the stage cost $\ell(x,u)= x_1^3$.
For $u=0$, the system has periodic solutions with period length $2\pi/(h\omega_0)=6$ and a numerical analysis reveals that the optimal period length (when allowing for controls $u$) is also $p^\star=6$, see Fig.~\ref{fig:orbit_lengths}.
The optimal orbit is
\begin{align*}
  \PiXstar &\approx \left(\left[\begin{smallmatrix}-1\\-0.5\end{smallmatrix}\right], \left[\begin{smallmatrix}0\\-1\end{smallmatrix}\right], \left[\begin{smallmatrix}0.7\\-0.5\end{smallmatrix}\right], \left[\begin{smallmatrix}0.7\\0.5\end{smallmatrix}\right], \left[\begin{smallmatrix}0\\1\end{smallmatrix}\right], \left[\begin{smallmatrix}-1\\0.5\end{smallmatrix}\right]\right).
\end{align*}
The corresponding inputs $\PiUstar$ are uniquely determined by $\PiXstar$ and the dynamics. 
Further, we can (approximately) verify the strict dissipativity Ass.~\ref{ass:diss} with respect to this orbit along the lines of \cite{Berberich2020}.

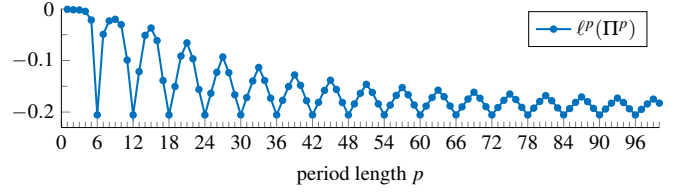
\begin{figure}
  \resizebox{\linewidth}{!}{\input{optimal_period_length}}
  \caption{The cost $\ell^p(\Pi^p)$ in Example~\ref{exmp:oscillator} of the optimal periodic orbit $\Pi^p$ of length $p$. We observe that the optimal period length is $p^\star = 6$ as $\ell^p(\Pi^p)$ is minimal for all $p\in 6\N$.}\label{fig:orbit_lengths}
\end{figure}

Concerning (i): When looking at closed-loop trajectories starting at $x_0^1=u_{\mathrm{max}} /\omega_0 \left[\begin{smallmatrix}-1 \\ -1\end{smallmatrix}\right]$ we observe that the undiscounted MPC scheme without terminal conditions starts the optimal open-loop trajectory with waiting at $x_0^1$ for all $N\geq 15$ with $[N]_{p^\star} = 3$.
For these horizon lengths, the system stays at $x_0^1$ for all times in closed loop, which is, however, far from being optimal.
Moreover, the MPC scheme fails not only for this specific initial condition, but also for a whole region of initial conditions around $x_0^1$. 
For these initial conditions, e.g., $x_0^2 = \left[\begin{smallmatrix} 0.1 \\ 0\end{smallmatrix}\right]$, the closed-loop trajectories are trapped in a $6$-periodic orbit around $x_0^1$ - again with a bad performance.
Only when initializing the state further away from $x_0^1$, e.g., at $x_0^3 = 2x_0^2 - x_0^1$ we observe that the undiscounted MPC $\mu_N^1$ finds the optimal orbit.
Compare also Fig.\ref{fig:ss}, where these observations are shown exemplary for $N=27$.
As $x_0^1$ is a problematic point, we take this initial condition also for the following performance analysis.
\begin{figure}
  \centering
  \resizebox{0.75\linewidth}{!}{\input{ss}}
  \caption{Closed-loop trajectories $x_{\mu_N^1}(\cdot, x_0)$ in Example~\ref{exmp:oscillator} of the undiscounted MPC scheme for different initial conditions $x_0\in \{x_0^1, x_0^2, x_0^3\}$.}\label{fig:ss}
\end{figure}
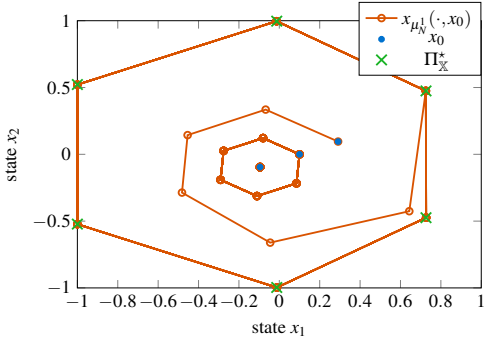

Concerning (ii): We compare the asymptotic average performance of the proposed LDE-MPC $\mu_N^\beta$ with a $1$-step MPC $\mu_N^1$ and a $p^\star$-step MPC $\nu_N$ as proposed by \cite{Mueller2016}. 
We simulate each scheme for different horizon lengths $N$ starting at $x_0$ for $T_\mathrm{sim} = 60$ time steps.
In most of the cases, an asymptotic $6$-periodic closed-loop behavior is reached after $T_\mathrm{sim}$ such that we can compute the asymptotic average performance by the average cost $J_\infty ^\mathrm{av}(x_0^1, \mu)$ of the last $6$ time steps for each MPC scheme $\mu\in\{\mu_N^\beta,\mu_N^1, \nu_N\}$.
As we can see in Fig.~\ref{fig:aap}, only the $p^\star$-step MPC $\nu_N$ and LDE-MPC $\mu_N^\beta$  achieve optimal performance (up to numerical accuracy of the simulation) for all horizons $N\geq 11$ and $N\ge15$, respectively.
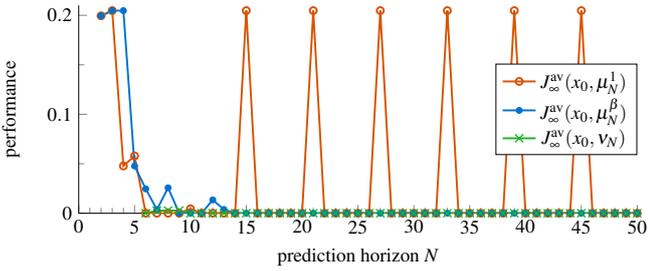
\begin{figure}
  \resizebox{\linewidth}{!}{\input{aap}}
  \caption{Asymptotic average performance in Example~\ref{exmp:oscillator} of the undiscounted MPC $\mu_N^1$, the linearly discounted MPC $\mu_N^\beta$, and the $p^\star$-step MPC $\nu_N$.}\label{fig:aap}
\end{figure}

Concerning (iii): 
Since the asymptotic average performance does not give any information about the transient cost to approach the asymptotic operating behavior, we compute the accumulated cost $J_T(x_0^1, \mu)-T\ell^\star$ for the three different MPC schemes $\mu\in\{\mu_N^\beta,\mu_N^1, \nu_N\}$. 
As this value depends very much on $[T]_{p^\star}$, we look at the average over all $T\in\I_{[25,30]}$ (one period): $J^\mathrm{tr}(x_0^1, \mu) = \frac 1 6 \sum_{T=25}^{30} J_T(x_0^1, \mu)-T\ell^\star$.
As we can observe in Fig.~\ref{fig:tp}, for horizons $N\geq 14$ LDE-MPC $\mu_N^\beta$ has the best transient performance of the three schemes.
The undiscounted MPC $\mu_N^1$ cannot reach this transient performance for any horizon length, whereas the $p^\star$-step MPC $\nu_N$ can reach it only for $[N]_{p^\star}=1$ and almost for $[N]_{p^\star}=0$.
This significant transient performance loss compared to $\mu_N^\beta$ can be explained by the fact that $\nu_N$ still executes the unrewarding first steps from $\mu_N^1$ before it converges to the optimal orbit, whereas $\mu_N^\beta$ directly goes in the right direction. 
\begin{figure}
  \resizebox{\linewidth}{!}{\input{tp}}
  \caption{Transient performance in Example~\ref{exmp:oscillator} of the undiscounted $\mu_N^1$, discounted $\mu_N^\beta$, and $p^\star$-step $\nu_N$ MPC scheme.}\label{fig:tp}
\end{figure}
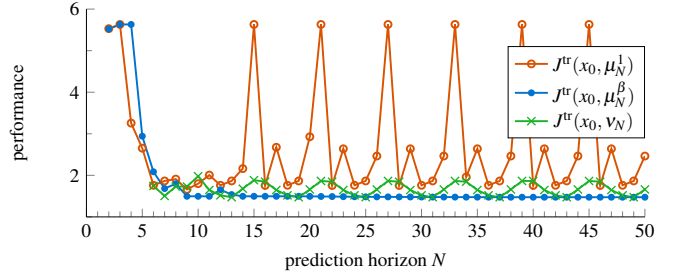

\end{exmp}

\begin{exmp}[Economic growth]\label{exmp:economic_growth}\ \\
  Consider the system $x(t+1)=u(t)$ with the cost $\ell(x,u) = -\log(5x^{0.34}-u)$.
  This system is a simple model for economic growth from~\cite{Brock1972} that has often been used as example in economic MPC, e.g., in \cite{Gruene2014}, \cite{Schwenkel2020b}, or \cite{Gruene2022}.
  Hence, it is well-known that this system is optimally operated at the steady state $\Pistar \approx (2.23, 2.23)$, i.e., $p^\star =1$.
  In \cite{Gruene2014} it is shown that the asymptotic average performance of an undiscounted MPC scheme without terminal conditions converges exponentially with $N\to\infty$ to the optimal performance $\ell^\star$. 
  Unfortunately, when introducing the linear discounts, this exponential convergence speed cannot be recovered and instead, the asymptotic average performance of LDE-MPC converges approximately proportional to $\sim \frac{1}{N^2}$ as we can estimate from the slope $\approx -2$ of the logarithmic plot of $J_\infty^\mathrm{av}(x_0,\mu_N^\beta)$ over $N$ in Fig.~\ref{fig:conv_speed}.
  The difference is huge in this example, e.g., in order to achieve optimality up to an error in the magnitude of $10^{-9}$ the undiscounted scheme needs a prediction horizon length of $N=9$, whereas the linearly discounted scheme needs $N=10^4$.
  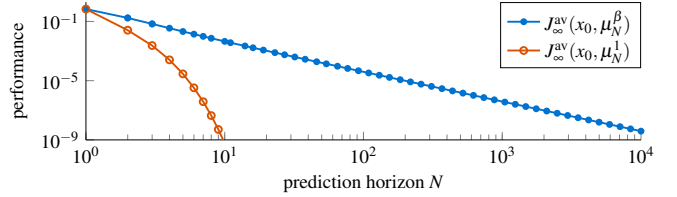
\begin{figure}
    \resizebox{\linewidth}{!}{\input{conv_speed}}
    \caption{Asymptotic average performance in Example~\ref{exmp:economic_growth} of the undiscounted $\mu_N^1$ and the discounted $\mu_N^\beta$ MPC scheme.}\label{fig:conv_speed}
  \end{figure}
\end{exmp}

%\begin{rem}
%  Interestingly, in these two examples and all other system/cost combinations we tested, the turnpike property holds not only in the weak formulation from \new{Definition~\ref{defn:wtp}}, but also in its standard strong formulation where $N-Q_\varepsilon(N,x)$ is bounded by a constant independent of $N$ instead of a constant times $\sqrt{N}$.
%\end{rem}

%% file: optimal_period_length.tex
\definecolor{mycolor1}{rgb}{0.00000,0.44700,0.74100}%
\begin{tikzpicture}

\begin{axis}[%
width=4in,
height=0.8in,
at={(1.46in,0.264in)},
scale only axis,
ytick = {-0.2, -0.1, 0},
minor y tick num=1,
xtick = {0,6, 12, 18, 24, 30, 36, 42, 48, 54, 60, 66, 72, 78, 84, 90, 96, 102},
minor x tick num=5,
xmin=0,
xmax=100,
ymin=-0.23,
axis x line*=bottom,
axis y line*=left,
ymax=0,
xlabel={period length $p$},
axis background/.style={fill=white}
]
\addplot [color=mycolor1, line width=1.0pt, mark size=1.3pt, mark=*, mark options={solid, mycolor1}]
  table[row sep=crcr]{%
    1	-0.000870791681846556\\
    2	-0.00174158336369308\\
    3	-0.00217697920916967\\
    4	-0.00478935428195642\\
    5	-0.0215435360731819\\
    6	-0.205615697834115\\
    7	-0.0492939799904643\\
    8	-0.0232853016184556\\
    9	-0.020224692724507\\
    10	-0.0305143234355635\\
    11	-0.0992176081512181\\
    12	-0.205615697834115\\
    13	-0.12150896802622\\
    14	-0.0513331632564379\\
    15	-0.037086873662942\\
    16	-0.0614258867034399\\
    17	-0.138813196627663\\
    18	-0.205615697834115\\
    19	-0.150414515045399\\
    20	-0.0914446867160536\\
    21	-0.0658240882012769\\
    22	-0.0968233221872075\\
    23	-0.156240045885299\\
    24	-0.205615697834115\\
    25	-0.163663140368429\\
    26	-0.122784342319798\\
    27	-0.0931733544874106\\
    28	-0.123584274950142\\
    29	-0.166455698010625\\
    30	-0.205615697834115\\
    31	-0.171782998636869\\
    32	-0.139884230410024\\
    33	-0.113517081732935\\
    34	-0.138813196627631\\
    35	-0.173168840837895\\
    36	-0.205615697834115\\
    37	-0.177269382290924\\
    38	-0.15041451504535\\
    39	-0.127952783586683\\
    40	-0.148193764956313\\
    41	-0.17791716137338\\
    42	-0.205615697834115\\
    43	-0.18122468213385\\
    44	-0.157942143069681\\
    45	-0.138307838819674\\
    46	-0.156240045885298\\
    47	-0.181453144751757\\
    48	-0.205615697834115\\
    49	-0.184211337117724\\
    50	-0.163662974872544\\
    51	-0.146226410468432\\
    52	-0.161937187922738\\
    53	-0.18418852811868\\
    54	-0.205615697834115\\
    55	-0.186546358285944\\
    56	-0.167749013405487\\
    57	-0.152477914401662\\
    58	-0.166455654464789\\
    59	-0.186367562326407\\
    60	-0.205615697834115\\
    61	-0.18842203102925\\
    62	-0.171782998559724\\
    63	-0.157538655680943\\
    64	-0.170126947994461\\
    65	-0.188144313296027\\
    66	-0.205615697834115\\
    67	-0.189961762384751\\
    68	-0.174768236801596\\
    69	-0.161556284815395\\
    70	-0.17316884083743\\
    71	-0.189620768328233\\
    72	-0.205615697834115\\
    73	-0.191248387216252\\
    74	-0.177191401910498\\
    75	-0.165079159708108\\
    76	-0.175730434810278\\
    77	-0.190867126471916\\
    78	-0.205615697834115\\
    79	-0.192339575364384\\
    80	-0.179395355954861\\
    81	-0.168080127209307\\
    82	-0.17791716137386\\
    83	-0.191933288257173\\
    84	-0.205615697834116\\
    85	-0.193276713420811\\
    86	-0.181224682133321\\
    87	-0.170667168158616\\
    88	-0.179805669250317\\
    89	-0.192855697889888\\
    90	-0.205615697834116\\
    91	-0.194090272832561\\
    92	-0.182815400549597\\
    93	-0.172914809967189\\
    94	-0.181453144751926\\
    95	-0.193661592624659\\
    96	-0.205615697834115\\
    97	-0.19480318571913\\
    98	-0.184211337114679\\
    99	-0.174895545393688\\
    100	-0.182646924679612\\
};
\addlegendentry{$\ell^p(\Pi^p)$};
\end{axis}
\end{tikzpicture}%

%% file: ss.tex
% !TeX root = lin_disc_empc.tex

\definecolor{mycolor1}{rgb}{0,0.447,0.81}%
\definecolor{mycolor2}{rgb}{0.85000,0.32500,0}%
\definecolor{mycolor3}{rgb}{0.1,0.7,0.15}%
\begin{tikzpicture}

\begin{axis}[%
width=3in,
height=2in,
at={(1.02in,0.92in)},
scale only axis,
xmin=-1,
xmax=1,
clip=false,
ymin=-1,
ymax=1,
xlabel = {state $x_1$},
ylabel = {state $x_2$},
legend style={at={(1,1.07)},anchor=north east},
axis background/.style={fill=white}
]
\addplot [color=mycolor2, line width=1.0pt, mark=o, mark size = 1.8pt, mark options={solid, mycolor2}]
  table[row sep=crcr]{%
-0.0954929658551372	-0.0954929658551372\\
-0.0954929788885295	-0.0954929623599343\\
-0.0954929884321601	-0.0954929718995816\\
-0.0954929849423986	-0.0954929849344319\\
-0.0954929719090063	-0.0954929884296348\\
-0.0954929623653757	-0.0954929788899875\\
-0.0954929658551373	-0.0954929658551373\\
-0.0954929788885295	-0.0954929623599343\\
-0.0954929884321601	-0.0954929718995816\\
-0.0954929849423986	-0.0954929849344319\\
-0.0954929719090063	-0.0954929884296348\\
-0.0954929623653757	-0.0954929788899875\\
-0.0954929658551373	-0.0954929658551373\\
-0.0954929788885295	-0.0954929623599343\\
-0.0954929884321601	-0.0954929718995816\\
-0.0954929849423986	-0.0954929849344319\\
-0.0954929719090063	-0.0954929884296348\\
-0.0954929623653757	-0.0954929788899875\\
-0.0954929658551373	-0.0954929658551373\\
-0.0954929788885295	-0.0954929623599343\\
-0.0954929884321601	-0.0954929718995816\\
-0.0954929849423986	-0.0954929849344319\\
-0.0954929719090063	-0.0954929884296348\\
-0.0954929623653757	-0.0954929788899875\\
-0.0954929658551373	-0.0954929658551373\\
-0.0954929788885295	-0.0954929623599343\\
-0.0954929884321601	-0.0954929718995816\\
-0.0954929849423986	-0.0954929849344319\\
-0.0954929719090063	-0.0954929884296348\\
-0.0954929623653757	-0.0954929788899875\\
-0.0954929658551373	-0.0954929658551373\\
-0.0954929788885295	-0.0954929623599343\\
-0.0954929884321601	-0.0954929718995816\\
-0.0954929849423986	-0.0954929849344319\\
-0.0954929719090063	-0.0954929884296348\\
-0.0954929623653757	-0.0954929788899875\\
-0.0954929658551373	-0.0954929658551373\\
-0.0954929788885295	-0.0954929623599343\\
-0.0954929884321601	-0.0954929718995816\\
-0.0954929849423986	-0.0954929849344319\\
-0.0954929719090063	-0.0954929884296348\\
-0.0954929623653757	-0.0954929788899875\\
-0.0954929658551373	-0.0954929658551373\\
-0.0954929788885295	-0.0954929623599343\\
-0.0954929884321601	-0.0954929718995816\\
-0.0954929849423986	-0.0954929849344319\\
-0.0954929719090063	-0.0954929884296348\\
-0.0954929623653757	-0.0954929788899875\\
-0.0954929658551373	-0.0954929658551373\\
-0.0954929788885295	-0.0954929623599343\\
-0.0954929884321601	-0.0954929718995816\\
-0.0954929849423986	-0.0954929849344319\\
-0.0954929719090063	-0.0954929884296348\\
-0.0954929623653757	-0.0954929788899875\\
-0.0954929658551373	-0.0954929658551373\\
-0.0954929788885295	-0.0954929623599343\\
-0.0954929884321601	-0.0954929718995816\\
-0.0954929849423986	-0.0954929849344319\\
-0.0954929719090063	-0.0954929884296348\\
-0.0954929623653757	-0.0954929788899875\\
-0.0954929658551373	-0.0954929658551373\\
};
\addlegendentry{$x_{\mu_N^1}(\cdot, x_0)$};
\addplot [color=mycolor2, line width=1.0pt, mark=o, mark size = 1.8pt, forget plot]
  table[row sep=crcr]{%
0.1	0\\
-0.0804458302734024	0.121555395261514\\
-0.275938805668393	0.026062419870627\\
-0.290985950798762	-0.190985950793758\\
-0.110540120518165	-0.312541346062627\\
0.0849528548984648	-0.217048370680975\\
0.100000000036946	7.59735332597189e-12\\
-0.080445830261509	0.121555395297309\\
-0.275938805668393	0.026062419870627\\
-0.290985950798762	-0.190985950793758\\
-0.110540120518165	-0.312541346062627\\
0.0849528548984648	-0.217048370680975\\
0.100000000036946	7.59735332597189e-12\\
-0.080445830261509	0.121555395297309\\
-0.275938805668393	0.026062419870627\\
-0.290985950798762	-0.190985950793758\\
-0.110540120518165	-0.312541346062627\\
0.0849528548984648	-0.217048370680975\\
0.100000000036946	7.59735332597189e-12\\
-0.080445830261509	0.121555395297309\\
-0.275938805668393	0.026062419870627\\
-0.290985950798762	-0.190985950793758\\
-0.110540120518165	-0.312541346062627\\
0.0849528548984648	-0.217048370680975\\
0.100000000036946	7.59735332597189e-12\\
-0.080445830261509	0.121555395297309\\
-0.275938805668393	0.026062419870627\\
-0.290985950798762	-0.190985950793758\\
-0.110540120518165	-0.312541346062627\\
0.0849528548984648	-0.217048370680975\\
0.100000000036946	7.59735332597189e-12\\
-0.080445830261509	0.121555395297309\\
-0.275938805668393	0.026062419870627\\
-0.290985950798762	-0.190985950793758\\
-0.110540120518165	-0.312541346062627\\
0.0849528548984648	-0.217048370680975\\
0.100000000036946	7.59735332597189e-12\\
-0.080445830261509	0.121555395297309\\
-0.275938805668393	0.026062419870627\\
-0.290985950798762	-0.190985950793758\\
-0.110540120518165	-0.312541346062627\\
0.0849528548984648	-0.217048370680975\\
0.100000000036946	7.59735332597189e-12\\
-0.080445830261509	0.121555395297309\\
-0.275938805668393	0.026062419870627\\
-0.290985950798762	-0.190985950793758\\
-0.110540120518165	-0.312541346062627\\
0.0849528548984648	-0.217048370680975\\
0.100000000036946	7.59735332597189e-12\\
-0.080445830261509	0.121555395297309\\
-0.275938805668393	0.026062419870627\\
-0.290985950798762	-0.190985950793758\\
-0.110540120518165	-0.312541346062627\\
0.0849528548984648	-0.217048370680975\\
0.100000000036946	7.59735332597189e-12\\
-0.080445830261509	0.121555395297309\\
-0.275938805668393	0.026062419870627\\
-0.290985950798762	-0.190985950793758\\
-0.110540120518165	-0.312541346062627\\
0.0849528548984648	-0.217048370680975\\
0.100000000036946	7.59735332597189e-12\\
};
\addplot [color=mycolor2, line width=1.0pt,  mark=o, mark size = 1.8pt, forget plot]
  table[row sep=crcr]{%
    0.290985931710274	0.0954929658551372\\
    -0.067652198732313	0.334700546812023\\
    -0.454131105831987	0.143714605563793\\
    -0.481971882488982	-0.286478916650766\\
    -0.0451689969650485	-0.661071823063205\\
    0.643615374000335	-0.425824434582221\\
    0.725535319780921	0.47492087713042\\
    -0.0135730300791335	0.996238286922517\\
    -1.00000000999565	0.521317409472594\\
    -1.00000000999357	-0.523117350707059\\
    -0.0120143112164586	-0.997138301377109\\
    0.727094114196855	-0.474020994536609\\
    0.725535395282793	0.474920833748956\\
    -0.0135729547105095	0.996238330532852\\
    -1.00000000999591	0.521317496505641\\
    -1.00000000999357	-0.523117263674093\\
    -0.0120143865965643	-0.997138257864867\\
    0.727094038824096	-0.474021038061574\\
    0.725535470662888	0.474920790228232\\
    -0.0135728793229819	0.996238373969553\\
    -1.00000000999582	0.521317583551151\\
    -1.00000000999357	-0.523117176628581\\
    -0.0120144619875412	-0.997138214346356\\
    0.727093963440471	-0.47402108159282\\
    0.725535546053845	0.474920746701231\\
    -0.0135728039170104	0.996238417660773\\
    -1.00000001003782	0.521317670710793\\
    -1.00000000999357	-0.523117089481115\\
    -0.0120145374684365	-0.997138170779661\\
    0.727093887970159	-0.474021125177845\\
    0.725535621532424	0.474920703122243\\
    -0.0135727284661159	0.996238461133425\\
    -1.0000000099954	0.521317757744228\\
    -1.00000000999357	-0.523117002424913\\
    -0.0120146128670577	-0.997138127253017\\
    0.727093812575669	-0.474021168711645\\
    0.725535696931494	0.47492065959085\\
    -0.0135726529193614	0.996238504665307\\
    -1.00000000999127	0.521317844976475\\
    -1.00000000999357	-0.523116915201991\\
    -0.0120146884133019	-0.997138083648595\\
};
\addplot [color=mycolor1, only marks, line width=1.0pt, mark size=1.3pt, mark=*]
table[row sep=crcr]{%
-0.0954929658551372	-0.0954929658551372\\
};
\addlegendentry{$x_0$};
\addplot [color=mycolor3, only marks, mark=x, mark size = 4pt, line width=1pt]
table[row sep=crcr]{%
-1.00000000999359	-0.52221737991059\\
-0.0127936328231522	-0.996688272122284\\
0.726314716630242	-0.474470892211694\\
0.726314716630242	0.474470892211694\\
-0.0127936328231524	0.996688272122284\\
-1.00000000999359	0.52221737991059\\
};
\addlegendentry{$\PiXstar$}
\addplot [color=mycolor1, line width=1.0pt, mark size=1.3pt, mark=*, mark options={solid, mycolor1}]
table[row sep=crcr]{%
  0.1	0\\
};
\addplot [color=mycolor1, line width=1.0pt, mark size=1.3pt, mark=*, mark options={solid, mycolor1}]
table[row sep=crcr]{%
  0.290985931710274	0.0954929658551372\\
};
\end{axis}
\end{tikzpicture}%

%% file: aap.tex
% !TeX root = lin_disc_empc.tex
\definecolor{mycolor1}{rgb}{0,0.447,0.81}%
\definecolor{mycolor2}{rgb}{0.85000,0.32500,0}%
\definecolor{mycolor3}{rgb}{0.1,0.7,0.15}%
\begin{tikzpicture}

\begin{axis}[%
width=4in,
height=1.5in,
at={(1.46in,0.51in)},
scale only axis,
xmin=0,
xmax=50,
ymin=0,
ytick = {0, 0.1, 0.2},
minor y tick num=1,
minor x tick num=4,
ymax=0.21,
axis background/.style={fill=white},
axis x line*=bottom,
axis y line*=left,
ylabel={performance},
xlabel={prediction horizon $N$},
legend style={at={(1,0.5)},anchor=east}
]
\addplot [color=mycolor2, line width=1.0pt, mark=o, mark size = 1.8pt, mark options={solid, mycolor2}]
  table[row sep=crcr]{%
2	0.199480227265211\\
3	0.204744906161363\\
4	0.0477655998312956\\
5	0.0579308076933009\\
6	4.43997061339019e-11\\
7	1.52567755281052e-05\\
8	8.17159302307635e-06\\
9	2.347717115464e-05\\
10	0.00451726833289975\\
11	2.34768691942111e-05\\
12	5.6486482158391e-12\\
13	9.5394872556831e-09\\
14	1.06088422347494e-05\\
15	0.20474490614334\\
16	2.34777028324806e-05\\
17	-1.94108062956388e-11\\
18	4.6715409318665e-13\\
19	1.08526753006544e-06\\
20	3.77167451071081e-06\\
21	0.204744906143336\\
22	2.34768820528142e-05\\
23	-2.16883733195061e-11\\
24	-1.7508772209851e-12\\
25	1.11054654361453e-07\\
26	2.34764415477651e-05\\
27	0.204744906143336\\
28	2.34768740336455e-05\\
29	-1.06968323088097e-11\\
30	-1.45186640487793e-12\\
31	7.92031698837459e-08\\
32	2.34764350211525e-05\\
33	0.204744906143336\\
34	2.34768745869252e-05\\
35	-1.84194326457998e-11\\
36	-1.39280253996787e-12\\
37	5.72121791886371e-08\\
38	2.34764284811617e-05\\
39	0.204744906143336\\
40	2.34768691456944e-05\\
41	-1.64108726607992e-11\\
42	4.30928959183063e-10\\
43	2.65573964763899e-08\\
44	2.34764150889022e-05\\
45	0.204744906143336\\
46	2.34769250080924e-05\\
47	-1.50359724671034e-11\\
48	2.93503749082547e-10\\
49	3.73251671237362e-09\\
50	2.34764039916957e-05\\
};
\addlegendentry{$J_\infty^\mathrm{av}(x_0, \mu_N^1)$};
\addplot [color=mycolor1, line width=1.0pt, mark size=1.3pt, mark=*, mark options={solid, mycolor1}]
table[row sep=crcr]{%
  2	0.199469642889747\\
  3	0.204744906170454\\
  4	0.204744906167424\\
  5	0.0477470652924176\\
  6	0.0246434003306534\\
  7	0.00365156533899264\\
  8	0.0257312710123661\\
  9	2.34772695492658e-05\\
  10	2.34768842141964e-05\\
  11	2.34776383632451e-05\\
  12	0.0134839107942328\\
  13	0.00381305430354548\\
  14	2.34768736668833e-05\\
  15	2.34768645052674e-05\\
  16	2.34770717617294e-05\\
  17	2.34768635356264e-05\\
  18	2.34768629568949e-05\\
  19	2.34768623902093e-05\\
  20	2.34768513751593e-05\\
  21	2.34768617785042e-05\\
  22	2.34768614079672e-05\\
  23	2.34768617206338e-05\\
  24	2.34768624255977e-05\\
  25	2.34768621223125e-05\\
  26	2.34768612170089e-05\\
  27	2.34768614844061e-05\\
  28	2.34768612246694e-05\\
  29	2.34768614882641e-05\\
  30	2.34768613348313e-05\\
  31	2.3476861391758e-05\\
  32	2.34768675154706e-05\\
  33	2.34768691659004e-05\\
  34	2.34768629294724e-05\\
  35	2.34768644207795e-05\\
  36	2.34768662862317e-05\\
  37	2.34768640250127e-05\\
  38	2.34785700404805e-05\\
  39	2.34768626060089e-05\\
  40	2.34768568812826e-05\\
  41	2.34768655016371e-05\\
  42	2.34768678339659e-05\\
  43	2.34768591692303e-05\\
  44	2.34768653963879e-05\\
  45	2.34768708521904e-05\\
  46	2.3476869447564e-05\\
  47	2.34768704229227e-05\\
  48	2.34768688487375e-05\\
  49	2.34768683836095e-05\\
  50	2.34768580301692e-05\\
};
\addlegendentry{$J_\infty^\mathrm{av}(x_0, \mu_N^\beta)$};
\addplot [color=mycolor3, line width=1.0pt, mark size=3pt, mark=x, mark options={solid, mycolor3, line width=0.7pt}]
  table[row sep=crcr]{%
6	2.34771137987699e-05\\
7	0.00262347146416159\\
8	0.00274630434263909\\
9	0.00271838152586332\\
10	5.87139781238477e-11\\
11	1.40278322535714e-05\\
12	2.34768576073408e-05\\
13	2.34768628261661e-05\\
14	2.34768603836755e-05\\
15	2.34768570367139e-05\\
16	2.34768606831581e-05\\
17	2.34768561082621e-05\\
18	2.34768573979804e-05\\
19	2.34768536769292e-05\\
20	2.34768621801829e-05\\
21	2.34768629198689e-05\\
22	2.34768628405158e-05\\
23	2.34768581798273e-05\\
24	2.34768622332515e-05\\
25	2.34768600415602e-05\\
26	2.34768597411061e-05\\
27	2.34768636903915e-05\\
28	2.34768611639402e-05\\
29	2.3476858446364e-05\\
30	2.34768614530978e-05\\
31	2.34768609888025e-05\\
32	2.34768613501524e-05\\
33	2.34768647974781e-05\\
34	2.34768616442504e-05\\
35	2.34768579457367e-05\\
36	2.34768609111979e-05\\
37	2.34768581964806e-05\\
38	2.34768594282175e-05\\
39	2.3476866019001e-05\\
40	2.34768660158924e-05\\
41	2.34768578023237e-05\\
42	2.34768696914522e-05\\
43	2.34768995921175e-05\\
44	2.34768694986065e-05\\
45	2.3476867318295e-05\\
46	2.34768620266945e-05\\
47	2.34768574361999e-05\\
48	2.34768564792431e-05\\
49	2.34768662215057e-05\\
50	2.34768709281574e-05\\
};
\addlegendentry{$J_\infty^\mathrm{av}(x_0, \nu_N)$};
\end{axis}
\end{tikzpicture}%

%% file: tp.tex
% !TeX root = lin_disc_empc.tex

\definecolor{mycolor1}{rgb}{0,0.447,0.81}%
\definecolor{mycolor2}{rgb}{0.85000,0.32500,0}%
\definecolor{mycolor3}{rgb}{0.1,0.7,0.15}%

\begin{tikzpicture}

\begin{axis}[%
width=4in,
height=1.5in,
at={(1.46in,0.974in)},
scale only axis,
xmin=0,
xmax=50,
ymin=1,
ymax=6,
axis background/.style={fill=white},
axis x line*=bottom,
axis y line*=left,
minor y tick num=1,
minor x tick num=4,
xlabel={prediction horizon $N$},
ylabel={performance},
legend style={at={(1,0.6)},anchor=east}
]
\addplot [color=mycolor2, line width=1.0pt, mark=o, mark size = 1.8pt, mark options={solid, mycolor2}]
  table[row sep=crcr]{%
    2	5.52895539672582\\
    3	5.63048491934187\\
    4	3.25685268659539\\
    5	2.65334475867416\\
    6	1.76055941595265\\
    7	1.86536752513243\\
    8	1.90987561527185\\
    9	1.6618580312126\\
    10	1.79979227351398\\
    11	2.00644165912042\\
    12	1.76055856583522\\
    13	1.86516524428822\\
    14	2.16288134296314\\
    15	5.63048491884634\\
    16	1.75628279217897\\
    17	2.6758234747334\\
    18	1.76055813802476\\
    19	1.86519186697761\\
    20	2.9283886200568\\
    21	5.63048491884629\\
    22	1.75628277831713\\
    23	2.63905863696755\\
    24	1.760557949852\\
    25	1.8651897845683\\
    26	2.4640951436979\\
    27	5.63048491884629\\
    28	1.75628277821199\\
    29	2.63905840489712\\
    30	1.76055790470871\\
    31	1.86518935545059\\
    32	2.46409514327708\\
    33	5.63048491884629\\
    34	1.96100421011738\\
    35	2.63905843636753\\
    36	1.76055783468787\\
    37	1.8651890833474\\
    38	2.46409514273852\\
    39	5.63048491884629\\
    40	1.75628277824863\\
    41	2.63905834538955\\
    42	1.76055777053753\\
    43	1.86518882547511\\
    44	2.46409514204714\\
    45	5.63048491884629\\
    46	1.75628277843442\\
    47	2.63905814740997\\
    48	1.76055769803129\\
    49	1.86518336240646\\
    50	2.46409514032313\\
};
\addlegendentry{$J^\mathrm{tr}(x_0, \mu_N^1)$};
\addplot [color=mycolor1, line width=1.0pt, mark size=1.3pt, mark=*, mark options={solid, mycolor1}]
table[row sep=crcr]{%
  2	5.5236726821982\\
  3	5.6304849195782\\
  4	5.63048491950854\\
  5	2.94214657352079\\
  6	2.08575258706332\\
  7	1.68239312669478\\
  8	1.79713867661965\\
  9	1.49596899186644\\
  10	1.49581327806242\\
  11	1.49581328851004\\
  12	1.65298478671536\\
  13	1.53746105586512\\
  14	1.49595328931795\\
  15	1.49581327774835\\
  16	1.49581328717316\\
  17	1.49581327768014\\
  18	1.49581327769332\\
  19	1.49581327680432\\
  20	1.49411301289541\\
  21	1.49165530961021\\
  22	1.48946254564444\\
  23	1.48693477663695\\
  24	1.48461154632493\\
  25	1.48250506370135\\
  26	1.4815530839151\\
  27	1.48152332694085\\
  28	1.48046309639761\\
  29	1.47950297539363\\
  30	1.47863077201656\\
  31	1.4778360859869\\
  32	1.47711000849083\\
  33	1.47644487280528\\
  34	1.47583405266107\\
  35	1.47527179776485\\
  36	1.47475309799065\\
  37	1.47427357306439\\
  38	1.47382940272084\\
  39	1.47370653515737\\
  40	1.47370653471785\\
  41	1.47370653488002\\
  42	1.47370653496117\\
  43	1.47370653484027\\
  44	1.47370653469052\\
  45	1.47370653488595\\
  46	1.47370653485252\\
  47	1.47370653481738\\
  48	1.47370653474634\\
  49	1.47370653480788\\
  50	1.47370653455041\\
};
\addlegendentry{$J^\mathrm{tr}(x_0, \mu_N^\beta)$};
\addplot [color=mycolor3, line width=1.0pt, mark size=3pt, mark=x, mark options={solid, mycolor3, line width=0.7pt}]
  table[row sep=crcr]{%
    6	1.74469330448435\\
    7	1.50391636025155\\
    8	1.7307307161924\\
    9	1.73079627229284\\
    10	1.97822166922101\\
    11	1.64729160413226\\
    12	1.51904148005218\\
    13	1.4737065346\\
    14	1.67869938434469\\
    15	1.88342079407047\\
    16	1.84675300717262\\
    17	1.64717096758189\\
    18	1.51904148005629\\
    19	1.47370653453153\\
    20	1.66027382181307\\
    21	1.86499522705878\\
    22	1.846753007204\\
    23	1.64717096765696\\
    24	1.51904148012748\\
    25	1.47370653459983\\
    26	1.66027382179805\\
    27	1.86499522707516\\
    28	1.84675300719362\\
    29	1.64717096765168\\
    30	1.51904148012982\\
    31	1.47370653461236\\
    32	1.6602738218107\\
    33	1.86499522709463\\
    34	1.8467530072042\\
    35	1.64717096761876\\
    36	1.51904148013337\\
    37	1.47370653456253\\
    38	1.66027382179878\\
    39	1.86499522711346\\
    40	1.86499522711371\\
    41	1.64717096761155\\
    42	1.51904148022532\\
    43	1.4737065349868\\
    44	1.66027382191043\\
    45	1.86499522713263\\
    46	1.846753007256\\
    47	1.64717096760821\\
    48	1.5190414800817\\
    49	1.47370653469255\\
    50	1.66027382193179\\
};
\addlegendentry{$J^\mathrm{tr}(x_0, \nu_N)$};
\end{axis}
\end{tikzpicture}%

%% file: conv_speed.tex
% !TeX root = lin_disc_empc.tex
\definecolor{mycolor1}{rgb}{0,0.447,0.81}%
\definecolor{mycolor2}{rgb}{0.85000,0.32500,0}%
\begin{tikzpicture}

\begin{axis}[%
width=4in,
height=1in,
at={(0.758in,0.481in)},
scale only axis,
xmode=log,
xmin=1,
xmax=10000,
xminorticks=true,
ymode=log,
ymin=1e-9,
ymax=2,
yminorticks=true,
axis x line*=bottom,
axis y line*=left,
xlabel={prediction horizon $N$},
ylabel={performance},
legend style={at={(1,1)},anchor=north east},
axis background/.style={fill=white}
]
\addplot [color=mycolor1, line width=1.0pt, mark size=1.3pt, mark=*, mark options={solid, mycolor1}]
table[row sep=crcr]{%
  1	0.685458831821469\\
  2	0.172532129555068\\
  2	0.172532129555068\\
  3	0.068247055826915\\
  3	0.068247055826915\\
  4	0.0350720453964413\\
  5	0.0209843310438171\\
  6	0.0138690926168494\\
  7	0.0098228711263979\\
  8	0.0073151966542242\\
  10	0.00450519745037914\\
  11	0.00367239432813937\\
  14	0.00220229448082776\\
  16	0.0016642330756953\\
  19	0.00116343899056992\\
  23	0.000783622825109687\\
  27	0.000563468383898424\\
  32	0.000397887058666502\\
  38	0.000280217589978538\\
  46	0.00019001251987194\\
  54	0.000137269341203616\\
  65	9.43314563235376e-05\\
  77	6.69992128041752e-05\\
  92	4.67965647272806e-05\\
  109	3.32605037280942e-05\\
  130	2.33354219758564e-05\\
  154	1.66016250271461e-05\\
  184	1.16126108045123e-05\\
  218	8.26326086711404e-06\\
  260	5.80336797262326e-06\\
  309	4.10534725370226e-06\\
  368	2.89244561435531e-06\\
  438	2.04059404307344e-06\\
  521	1.44150143244204e-06\\
  620	1.01748174730076e-06\\
  737	7.19820585359798e-07\\
  877	5.08197524062837e-07\\
  1044	3.58528253308066e-07\\
  1242	2.53274423211991e-07\\
  1478	1.78817499252304e-07\\
  1758	1.26374004372209e-07\\
  2092	8.92316047718111e-08\\
  2489	6.30300163173558e-08\\
  2962	4.45030079454511e-08\\
  3524	3.14380788069002e-08\\
  4193	2.22050333675838e-08\\
  4989	1.568381691186e-08\\
  5937	1.10745430603743e-08\\
  7064	7.82244846675439e-09\\
  8404	5.52660894825863e-09\\
  10000	3.90318932730338e-09\\
};
\addlegendentry{$J_\infty^\mathrm{av}(x_0, \mu_N^\beta)$};

\addplot [color=mycolor2, line width=1.0pt, mark=o, mark size = 1.8pt, mark options={solid, mycolor2}]
table[row sep=crcr]{%
  1	0.685458831821469\\
  2	0.0256512326473008\\
  3	0.00235508659599493\\
  4	0.000253958328866766\\
  5	2.86956636026403e-05\\
  6	3.29189894876514e-06\\
  7	3.79557375529416e-07\\
  8	4.38381886347372e-08\\
  9	5.06617592321845e-09\\
  10	5.85589798873798e-10\\
  11	6.76916300790253e-11\\
  12	7.82529596676795e-12\\
  13	9.04609720464578e-13\\
  14	1.04805053524615e-13\\
  15	1.22124532708767e-14\\
};
\addlegendentry{$J_\infty^\mathrm{av}(x_0, \mu_N^1)$};
\end{axis}

\end{tikzpicture}%

%% file: conclusion.tex
\section{Discussion and conclusion}\label{sec:conclusion}

In this work, we have shown that LDE-MPC is recursively feasible when initialized at any $x_0\in\X_0$ and achieves an asymptotic average performance that is optimal up to any desired level of suboptimality when the prediction horizon is chosen sufficiently large.
Moreover, we have proven practical asymptotic stability and Example~\ref{exmp:oscillator} has revealed improvements in the transient performance compared to state-of-the-art methods.
One further clear advantage of LDE-MPC is that it does not need any offline design, neither terminal conditions need to be designed as in \cite{Zanon2017}, nor the optimal period length $p^\star$ must be computed as in \cite{Mueller2016}.

When facing real world applications, it is often impracticable to design terminal conditions and instead practitioners omit terminal conditions and increase the prediction horizon $N$ until the closed-loop behavior is satisfactory.
The work of \cite{Gruene2013} provides a theoretical justification for this procedure in the case where optimal operation is a steady state.
Similarly, in the case where optimal operation is a steady state \emph{or} a periodic orbit, the present work provides a theoretical justification to implement LDE-MPC and increase its prediction horizon $N$ until the desired performance is reached.

We identified various future research questions:
\begin{itemize}
\item Are also other discount functions possible? If other discount functions give the same qualitative performance and stability guarantees as LDE-MPC, then the choice of the discount functions may be exploited as design parameter to improve the quantitative performance.
Hence, a theoretical analysis is needed to find the properties a discount function must have to guarantee performance and stability as in Theorems~\ref{thm:performance} and~\ref{thm:stability} and then, a numerical analysis could give insights which discount has the best quantitative behavior.
In particular, it is desirable to find a discount that recovers the exponential convergence of undiscounted MPC (compare  Example~\ref{exmp:economic_growth}).
An educated guess therefor could be to start the linear discounting only in the second half of the prediction horizon and keep the first half undiscounted.
\item Does LDE-MPC also work with other optimal operating behaviors? 
In this work, we investigated LDE-MPC for setups that are optimally operated at a periodic orbit, however, the design does not use any information of the periodic optimal operating behavior. 
Hence there is justified hope that the results of this work may be generalized to other optimal operating behaviors, e.g., quasi-periodic or the ones considered in \cite{Dong2018a}, \cite{Martin2019}, or \cite{Gruene2019}. 
\item Can we give transient performance guarantees? We have already observed in Example~\ref{exmp:oscillator} that LDE-MPC has a comparably good transient performance. 
Based on the practical stability guarantee from Section~\ref{sec:stability}, we conjecture that one may obtain transient performance guarantees, similar to \cite{Gruene2014}, \cite{Gruene2015}, or \cite{Kloeppelt2021}.
\item \new{It may be possible to extend LDE-MPC to a setting with disturbances such that robust performance and robust stability guarantees can be given. 
Such extensions are known for undiscounted economic MPC in the case when the optimal operating behavior is periodic (see \cite{Wabersich2018}, \cite{Broomhead2015}) but terminal conditions are used, as well as when no terminal conditions are used but the optimal operating behavior is a steady state (see \cite{Schwenkel2020b}).}
\end{itemize}

%% file: main_proofs.tex
\section{Technical parts of the proofs}\label{sec:proofs}

This section contains all the technical parts of the proofs of Thm.~\ref{thm:wtp},~\ref{thm:performance}, and~\ref{thm:stability} in full detail.
Since these proofs are rather long and technical, we structured them into several Lemmas, which hopefully helps the reader understanding the idea of the proof much better.
In addition, we provide an overview of the whole proof structure in Fig.~\ref{fig:proof_struct}.

\newcommand{\rcf}{Rotated cost functional}
\newcommand{\wtplb}{Bounded rotated cost $\Rightarrow$ weak turnpikes}
\newcommand{\wtpub}{Uniform bound from reachability}
\newcommand{\costAtOrbit}{Cost of trajectories close to orbit}
\newcommand{\dpp}{Dynamic programming principle}
\newcommand{\VNdiff}{Almost optimal candidate solution}
\newcommand{\simcost}{Difference of value functions}
\newcommand{\rotCostZero}{Rot. stage cost is positive definite}
\newcommand{\rotValUB}{Rot. value function is positive definite}
\begin{figure}[tb]\centering
  \resizebox{\linewidth}{!}{
  \begin{tikzpicture}[node distance=0.4cm]
    \tikzset{lemmana/.style 2 args={
      draw, minimum width = 2.5cm, minimum height = 1.24cm,
        path picture = {
          \node [minimum width=0.1cm, minimum height=0cm,inner ysep=0.08cm,inner xsep=0.1cm, anchor=north west] at (path picture bounding box.north west) (tmpnode) {\fontsize{5}{4}\selectfont #1\phantom{,}};
          \draw[fill=white] (tmpnode.south west) -- ($(tmpnode.south east)+(-0.2,0)$)  to [out=0, in=-90] ($(tmpnode.south east)+(0,0.2)$) -- (tmpnode.north east) -| cycle;
          \node [minimum width=0.1cm, minimum height=0.1cm,inner sep=0.1cm, anchor=north west] at (path picture bounding box.north west) {\tiny #1\vphantom{,}};
          \node [minimum width=0.1cm, minimum height=0.1cm,inner sep=0.05cm, anchor=south] at (path picture bounding box.south) {\begin{minipage}{2.4cm}\centering\fontsize{5}{4}\selectfont #2\end{minipage}};
        }  
      }
    }
    \tikzset{
      lemman/.style 2 args={
        draw, minimum width = 2.5cm, minimum height = 1.24cm,
        path picture = {
          \node [minimum width=0.1cm, minimum height=0.1cm,inner sep=0.05cm, anchor=south] at (path picture bounding box.south) {\begin{minipage}{2.3cm}\centering\fontsize{5}{4}\selectfont #1\end{minipage}};
        }
      }
    }
    %%% LEMMA 23 %%%
    \node[lemman={\rcf}] (lem23) {Lemma~\ref{lem:rcf}};
    %%% LEMMA 25 %%%
    \node[lemmana={Ass.~\ref{ass:diss}}{\wtplb}, node distance=0.7cm, right=of lem23] (lem25) {Lemma~\ref{lem:wtp_lb}};
    %%% THEOREM 9 %%%
    \node[lemmana={Ass.~\ref{ass:diss},~\ref{ass:loc_ctrb}}{\wtp}, node distance=0.7cm, right=of lem25, fill=black!10] (thm9) {\textbf{Theorem~\ref{thm:wtp}}};
    \draw [line width=1pt](thm9.north west) -| (thm9.south east) -| cycle;
    %%% COROLLARY 10 %%%
    \node[lemmana={Ass.~\ref{ass:diss},~\ref{ass:reachability}}{\rf}, above=of thm9, fill=black!10] (cor10) {\textbf{Corollary~\ref{cor:rf}}};
    \draw [line width=1pt](cor10.north west) -| (cor10.south east) -| cycle;
    %%% LEMMA 24 %%%
    \node[lemmana={Ass.~\ref{ass:tech},~\ref{ass:diss}}{\wtpub}, above=of lem25] (lem24) {Lemma~\ref{lem:wtp_ub}};
    %%% LEMMA 29 %%%
    \node[lemmana={Ass.~\ref{ass:tech},~\ref{ass:diss},~\ref{ass:lambda_cont}}{\rotCostZero}, below =of lem23] (lem29) {Lemma~\ref{lem:rot_cost_zero}};
    %%% LEMMA 26 %%%
    \node[lemmana={Ass.~\ref{ass:tech},~\ref{ass:diss}}{\costAtOrbit}, below=of thm9] (lem26) {Lemma~\ref{lem:cost_of_traj_at_orbit}};
    %%% LEMMA 30 %%%
    \node[lemmana={Ass.~\ref{ass:tech},~\ref{ass:loc_ctrb}}{\rotValUB}, below=of lem29](lem30) {Lemma~\ref{lem:rot_val_ub}};
    %%% LEMMA 31 %%%
    \node[lemmana={\!\!\!\!\!{\begin{tabular}{l}\ \\[-0.25cm]Ass.~\ref{ass:tech},~\ref{ass:diss},~\ref{ass:loc_ctrb},~\ref{ass:lambda_cont}\\ $\Pistar$ minimal\\[0.08cm]\end{tabular}}\!\!\!\!\!\!}{\simcost}, below=of lem25] (lem31) {\raisebox{-0.4cm}{Lemma~\ref{lem:sim_cost}}};
    %%% LEMMA 27 %%%    
    \node [lemmana={Ass.~\ref{ass:tech},~\ref{ass:diss},~\ref{ass:loc_ctrb}}{\VNdiff}, below=of lem26] (lem27) {Lemma~\ref{lem:V_N_diff}};
    %%% THEOREM 15 %%%
    \node[lemman={\lyapunov}, below=of lem30] (thm15) {Theorem~\ref{thm:lyapunov}};
    %%% THEOREM 15 %%%
    \node[lemmana={Ass.~\ref{ass:diss}}{\stability}, below=of lem31, fill=black!10] (thm17) {\textbf{Theorem~\ref{thm:stability}}};
    \draw [line width=1pt](thm17.north west) -| (thm17.south east) -| cycle;
    %%% LEMMA 28 %%%
    \node [lemmana={Ass.~\ref{ass:tech}}{\dpp}, below=of thm17] (lem28) {Lemma~\ref{lem:dpp}};
    %%% THEOREM 11 %%%
    \node[lemmana={Ass.~\ref{ass:tech}}{\performance}, below=of lem27, fill=black!10] (thm11) {\textbf{Theorem~\ref{thm:performance}}};
    \draw [line width=1pt](thm11.north west) -| (thm11.south east) -| cycle;
    %%% THEOREM 15 %%%
    %%% ARROWS %%%
    \draw[-latex] (lem23.north east) to [bend left=14] (thm9.north west);
    \draw[-latex] (lem23.south east) -- (lem31.north west);
    \draw[-latex] (lem23.south west) to [bend right=14] (lem30.north west);
    \draw[-latex] (lem24) -- (cor10);
    \draw[-latex] (lem24.south east) -- (thm9.north west);
    \draw[-latex] (thm9) -- (cor10);
    \draw[-latex] (lem25) -- (thm9);
    \draw[-latex] (lem29) -- (lem30);
    \draw[-latex] (lem29) -- (lem31);
    \draw[-latex] (lem27.east) to [bend right=14] (thm9.south east);
    \draw[-latex] (thm9.east) to [bend left=14] (thm11.north east);
    \draw[-latex] (lem30.north east) -- (lem31.south west);
    \draw[-latex] (thm9.south west) -- (lem31.north east);
    \draw[-latex] (thm9.south west) -- (thm17.north east);
    \draw[-latex] (lem26) -- (lem27);
    \draw[-latex] (lem27) -- (thm11);
    \draw[-latex] (lem27) -- (thm17);
    \draw[-latex] (thm15.north east) -- (thm17.south west);
    \draw[-latex] (lem31) -- (thm17);
    \draw[-latex] (lem30) -- (thm17);
    \draw[-latex] (lem28) -- (thm17);
    \draw[-latex] (lem28) -- (thm11);
  \end{tikzpicture}}
  \caption{Overview of the structure of the proof of the main results of this article. Each proof uses the assumptions in the upper left corner and the Lemmas or Theorems pointing to it. 
   Obviously, their assumptions are then also required.}\label{fig:proof_struct}
\end{figure}
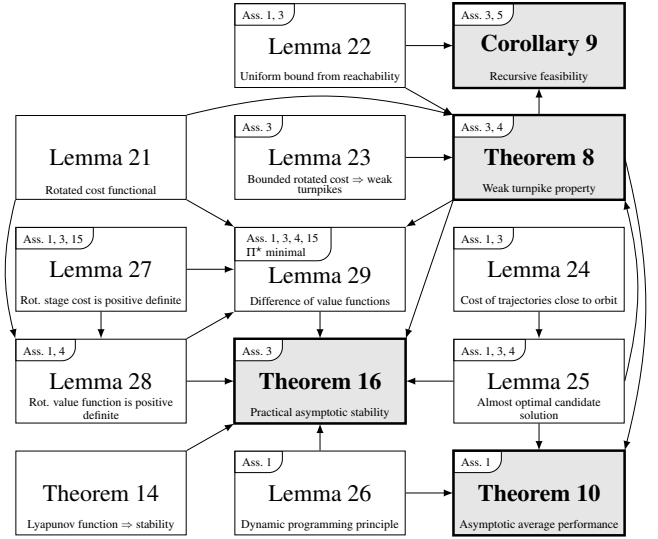

\begin{rem}\label{rem:wlog_ell_nn}
  Throughout this section, we will assume that $\ell$ is non-negative.
  This assumption is without loss of generality and therefore commonly made in undiscounted economic MPC to simplify the analysis.
  Let us briefly justify that this assumption still goes without loss of generality if we discount the cost function.
  We know due to Ass.~\ref{ass:tech} that $\ell$ is continuous and thus lower bounded on the compact set $\X\times \U$, i.e., $\ell_\mathrm{min}:=\inf_{(x,u)\in \X\times \U}\ell(x,u)$ exists and is finite.
  Thus, we can redefine the cost as $\bar \ell(x,u) := \ell(x,u)-\ell_\mathrm{min}\geq 0$, which implies that the resulting cost functional $\bar J_N^\beta$ for $x\in\X$ and $u\in\U(x)$ is 
  \begin{align*}
    \bar J_N^\beta(x,u)=J_N^\beta (x,u) - \ell_\mathrm{min}\sum_{k=0}^{N-1}\beta_N(k).
  \end{align*}
  Therefore, the cost functionals $\bar J_N^\beta$ and $J_N^\beta$ differ only by a constant and thus have the same minimizer $u_{N,x}^\beta$.
  Hence, the MPC control law $\mu_N^\beta $ as defined in \eqref{eq:mpc_cl} remains unchanged and we can indeed assume $\ell(x,u)\geq 0$ without loss of generality.
\end{rem}

\begin{lem}[\rcf]\label{lem:rcf}\ \\
  For all $N\in\N$, $x\in\X$ and $u\in\U^N(x)$ the rotated cost function satisfies
  \begin{align}
      \tilde J_{N}^\beta (x,u) &= J_{N}^\beta (x,u)-\new{\frac{N+1}{2}}\ell^\star  + \lambda (x) -\frac 1 N \sum_{k=1}^{N} \lambda(x_u(k,x)). \label{eq:rcf}
  \end{align}
\end{lem}
\begin{pf}
We compute 
\begin{align*}
  \tilde J_{N}^\beta (x,u)  &\refeq{\eqref{eq:rot_cost}}{=} J_{N}^\beta (x,u) - \sum_{k=0}^{N-1} \!\beta_N(k) \ell^\star \\
  &\quad + \sum_{k=0}^{N-1} \!\beta_N(k) \Big(\lambda\big(x_{u}(k,x)\big)- \lambda\big(x_{u}(k+1,x)\big)\!\Big)
\end{align*}
and \new{$\sum_{k=0}^{N-1} \!\beta_N(k) \ell^\star=\frac{N+1}{2}\ell^\star$ as well as}
\begin{align*}
  \sum_{k=0}^{N-1} &\beta_N(k) \Big(\lambda\big(x_{u}(k,x)\big)- \lambda\big(x_{u}(k+1,x)\big)\!\Big)\\
  &=\beta_N(0)\lambda(x)-\beta_N(N-1)\lambda\big(x_{u}(N,x)\big)\\&\quad  + \sum_{k=1}^{N-1} \!\lambda\big(x_{u}(k,x)\big) \big(\beta_N(k)- \beta_N(k-1)\big) \\
  &\refeq{\eqref{eq:beta}}{=} \lambda(x) -\frac 1 N \sum_{k=1}^{N} \lambda\big(x_{u}(k,x)\big).\\[-1.5cm]
\end{align*}
\hfill $\square$\\
\end{pf}

\new{
\begin{lem}[\wtpub]\label{lem:wtp_ub}\ \\
  Let Ass.~\ref{ass:tech} and~\ref{ass:diss} hold. If $\PiXstar$ can be reached from $x\in \X$ within $M$ steps, i.e., there exists $\bar u\in \U^M(x)$ with $x_{\bar u}(M,x)=\PiXstar(l)$ for some $l\in\I_{[0,p^\star -1]}$, then for all $N\in\N$
    \begin{align*}
       V_N^\beta(x)-\frac{N+1}{2}\ell^\star +\lambda(x)+\bar \lambda \leq C(M)
    \end{align*}
    holds with $C(M)=M(\ell_\mathrm{max}-\ell^\star)+2\bar \lambda +\ell^\star p^\star $.
\end{lem}
\begin{pf}
We derive this upper bound by
constructing a suitable suboptimal candidate solution for $V_{N}^\beta(x)$.
We can extend the input $\bar u$ to any horizon length $\bar u \in \U^N(x)$ by remaining on the optimal orbit, i.e., choose $\bar u(k):= \PiUstar([k-M+l]_{p^\star})$ for $k\in \I_{[M,N-1]}$
which results in the state $x_{\bar u}(k,x)=\PiXstar([k-M+l]_{p^\star})$ for all $k\in \I_{[M,N]}$.
Therefore, noting that $\sum_{k=0}^{N-1} \!\beta_N(k) \ell^\star=\frac{N+1}{2}\ell^\star$, we have
\begin{align*}
  &V_N^\beta(x)-\frac{N+1}{2}\ell^\star\leq J_N^\beta (x,\bar u) - \sum_{k=0}^{N-1} \beta_N(k) \ell^\star \\
  &= \underbrace{\sum_{k=0}^{M-1}  \beta_N(k) \Big(\ell(x_{\bar u}(k,x),\bar u(k)) - \ell^\star\Big)}_{\leq M(\ell_\mathrm{max}-\ell^\star)} \lambda\\
  &\qquad + \sum_{k=M}^{N-1}  \beta_N(k) \Big(\ell(\Pistar([k-M-l]_{p^\star})) - \ell^\star\Big)
\end{align*}
where we introduced $\ell_\mathrm{max}:=\sup_{(x,u)\in \X\times \U} \ell(x,u)$, which is finite due to Ass.~\ref{ass:tech}.
The last line can be upper bounded by using that $\ell$ is non-negative (compare Remark~\ref{rem:wlog_ell_nn}) and that $\beta_N$ is decreasing.
In particular, we know that 
\begin{align*}
  &\sum_{k=M}^{N-1}  \beta_N(k) \ell(\Pistar([k-M+l]_{p^\star})) \leq \hspace{-0.1cm} \sum_{j=0}^{\lfloor (N-M)/p^\star \rfloor} \hspace{-0.2cm}\beta_N(M+jp^\star) \ell^\star p^\star
\end{align*}
and that 
\begin{align*}
  -\sum_{k=M}^{N-1}  \beta_N(k) \ell^\star \leq - \sum_{j=1}^{\lfloor (N-M)/p^\star \rfloor} \beta_N(M+jp^\star)\ell^\star p^\star.
\end{align*}
Altogether and using $\beta_N(M)\leq 1$ and $|\lambda(x)|\leq \bar \lambda $ from Assumption~\ref{ass:diss}, this yields
\begin{align*}
  V_N^\beta(x)-&\frac{N+1}{2}\ell^\star +\lambda(x)+\bar \lambda \\&\leq M(\ell_\mathrm{max}-\ell^\star)+2\bar \lambda +\ell^\star p^\star = C(M).\quad  \square
\end{align*}
\end{pf}}

\new{
\begin{lem}[\wtplb]\label{lem:wtp_lb}
    Let Assumption~\ref{ass:diss} hold and let $x\in \X$. If there exist $C\in \R$, $N_0\in\N$ such that the rotated cost function $\tilde J^\beta_N(x,u_{N,x}^\beta)\leq C$ is bounded for all $N\geq N_0$, then the weak turnpike property is satisfied at $x$, i.e., $x\in\X_{\alpha,N_0}$ with $\alpha(\varepsilon)=\frac{1}{\sqrt{2C}}\sqrt{\underline\alpha_{\tilde \ell}(\varepsilon)}$.
\end{lem}
\begin{pf}
  In order to quantify how close the trajectory resulting from $u_{N,x}^\beta$ is to the optimal periodic orbit $\Pistar$, we sum up the dissipation inequality \eqref{eq:strict_diss} from Ass.~\ref{ass:diss} and obtain
  \begin{align*}
    &\sum_{k=0}^{N-1} \beta_N(k) \underline\alpha_{\tilde \ell}\left(\left\|\big(x_{u_{N,x}^\beta}(k,x),u_{N,x}^\beta (k)\big)\right\|_{\Pistar}\right)\nonumber \\
    &\leq \sum_{k=0}^{N-1} \beta_N(k) \tilde \ell\left(x_{u_{N,x}^\beta}(k,x),u_{N,x}^\beta (k)\right)=\tilde J_{N}^\beta (x,u_{N,x}^\beta)\leq C.
  \end{align*}
  Next, we lower bound the left hand side by taking only the $N-Q_\varepsilon(N,x)$ points outside the $\varepsilon$-neighborhood of $\Pistar$ and bound their norm by $\left\|\big(x_{u_{N,x}^\beta}(k,x),u_{N,x}^\beta (k)\big)\right\|_{\Pistar}\geq \varepsilon$.
  Since we do not know which ones out of the total $N$ points these are, we consider the worst case, i.e., that these points are the ones with the smallest weights $\beta_N(k)$, which are the last $N-Q_\varepsilon(N,x)$ points.
  Hence, by exploiting linearity of $\beta_N$, we can explicitly compute
  \begin{align*}
    \sum_{k=0}^{N-1} &\beta_N(k) \underline\alpha_{\tilde \ell}\left(\left\|\big(x_{u_{N,x}^\beta}(k,x),u_{N,x}^\beta (k)\big)\right\|_{\Pistar}\right) \\&\geq\sum_{k=Q_{\varepsilon}^{\beta}(N,x)}^{N-1} \beta_N(k)\underline\alpha_{\tilde \ell}(\varepsilon)\refeq{\eqref{eq:beta}}{=} \sum_{k=Q_{\varepsilon}^{\beta}(N,x)}^{N-1} \frac{N-k}{N}\underline\alpha_{\tilde \ell}(\varepsilon)\\
    &=\frac {\underline\alpha_{\tilde \ell}(\varepsilon)} N \sum_{k=1}^{N-Q_{\varepsilon}^{\beta}(N,x)} k  \\ 
    &= \frac{\big(N-Q_{\varepsilon}^{\beta}(N,x)\big)\big(N-Q_{\varepsilon}^{\beta}(N,x)+1\big)}{2N} \underline\alpha_{\tilde \ell}(\varepsilon)\\
    &\geq\frac{\big(N-Q_{\varepsilon}^{\beta}(N,x)\big)^2}{2N}\underline\alpha_{\tilde \ell}(\varepsilon),
  \end{align*}
  Putting these two pieces together yields
  \begin{align*}
    \frac{\big(N-Q_{\varepsilon}^{\beta}(N,x)\big)^2}{2N}\underline\alpha_{\tilde \ell}(\varepsilon) \leq C \ 
    \Leftrightarrow\  N-Q_{\varepsilon}^{\beta}(N,x) \leq \frac{\sqrt{2NC}}{\sqrt{\underline\alpha_{\tilde \ell}(\varepsilon)}}.
  \end{align*}
  After defining $\alpha\in\Kinf$ as $\alpha(\varepsilon):=\frac{1}{\sqrt{2C}}\sqrt{\underline\alpha_{\tilde \ell}(\varepsilon)}$, we obtain the desired result.\hfill $\square$
\end{pf}}

\begin{lem}[\costAtOrbit]\label{lem:cost_of_traj_at_orbit}
    Let Ass.~\ref{ass:tech} and~\ref{ass:diss} hold. Then, there exists $\alpha_1 \in \Kinf$ such that for all $\varepsilon>0$ and for all trajectories of length $T$ defined by $\hat x\in\X$ and $\hat u\in\U^{T}(\hat x)$ that satisfy $\|(x_{\hat u}(k,\hat x),\hat u(k))\|_\Pistar\leq \varepsilon$ for all points $k\in \I_{[0,T-1]}$, the following bound on the cost holds
  \begin{align}\label{eq:cost_of_traj_at_orbit}
    \sum_{k=0}^{T-1} \ell(x_{\hat u}(k,\hat x),\hat u(k)) &\leq (T+ p^\star-1) \ell^\star + T\alpha_1(\varepsilon).
  \end{align}
\end{lem}
\begin{pf}
This lemma is a consequence of the continuity of $f$ and $\ell$ and the fact that Ass.~\ref{ass:diss} implies that no closed orbit can have a better average performance than $\ell^\star$. 
Let us therefore formulate the continuity of $f$ and $\ell$ on the periodic orbit $\Pistar$ in terms of $\Kinf$ functions, i.e., there exist $\overline\alpha_f,\overline\alpha_\ell \in \Kinf$ such that for all $j\in\I_{[0,p^\star-1]}$ and all $(x,u)\in\X\times \U$
\begin{align*}
  \|f(x,u) - f(\Pistar(j))\|\leq \overline\alpha_f(\|(x,u)-\Pistar(j)\|)\\
  |\ell(x,u) - \ell(\Pistar(j))|\leq \overline\alpha_\ell(\|(x,u)-\Pistar(j)\|)
\end{align*}
holds.
Using the continuity of $\ell$ we immediately get
\begin{align}\label{eq:follow_orbit1}
  \sum_{k=0}^{T-1} \ell(x_{\hat u}(k,\hat x),\hat u(k)) &\leq \sum_{k=0}^{T-1} \ell(\Pistar(i_k)) + T \overline\alpha_\ell(\varepsilon),
\end{align}
where the sequence $(i_k)_{k\in \I_{[0, T -1]}}$ satisfies $\|(x_{\hat u}(k,\hat x),\hat u(k))-\Pistar(i_k)\|\leq \varepsilon$ for all $k \in \I_{[0, T -1]}$.
Thus, it remains to compute $\sum_{k=0}^{T-1} \ell(\Pistar(i_k))$.
To this end, let $\bar \varepsilon$ be the minimum distance between any two points on the periodic orbit $\PiXstar$ which are not equal. 
Thus, $\bar \varepsilon>0$, since there are only finitely many points and we excluded distance $0$, i.e., equal points. 
Now, choose $\varepsilon_1>0$ such that $\varepsilon_1+\overline\alpha_f(\varepsilon_1)<\bar \varepsilon$.
This choice guarantees for all $\varepsilon<\varepsilon_1$, all $(x,u)\in\X\times \U$ with $\|(x,u)-\Pistar(i_k)\|\leq \varepsilon$, and all $j$ with $\PiXstar (j) \neq f(\Pistar(i_k))=\PiXstar ([i_k+1]_{p^\star})$ that
\begin{align*}
  &\|f(x,u)-\PiXstar(j)\|\\&\geq \|f(\Pistar(i_k))-\PiXstar(j)\|-\|f(x,u)-f(\Pistar(i_k))\|\\& \geq \|\PiXstar([i_k+1]_{p^\star})-\PiXstar(j)\|-\overline\alpha_f(\|(x,u)-\Pistar(i_k)\|) \\& \geq \bar \varepsilon - \overline\alpha_f(\varepsilon)> \varepsilon.
\end{align*}
Hence, since $\|\PiXstar (i_{k+1}) -f (x_{\hat u}(k,\hat x),\hat u(k))\|\leq \varepsilon$ the only possibility that is left is
\begin{align}\label{eq:follow_orbit}
\PiXstar (i_{k+1}) = \PiXstar([i_k+1]_{p^\star}).
\end{align}
Further, let us denote the sequence that starts at $i_0$ and follows the orbit $\Pistar$ for $Tp^\star$ time steps with $\bar i$, i.e., $\bar i_k = [i_0+k]_{p^\star}$ for all $k\in [0, Tp^\star -1]$.
Obviously, $\bar i$ describes $T$ full orbits $\Pistar$, starting at the phase $i_0$, and thus, $\bar i$ has the average cost $\ell^\star$.
Now, we will transform $\bar i$ to $i$ by taking away parts of $\bar i$ that are an orbit themselves until we are left with $i$.
Thereby, we will see that the lower bound on the average cost always stays $\ell^\star$.
In particular, apply the following algorithm:
\begin{enumerate}[(i)]
  \item Find the smallest $k\in \I_{[0,T-1]}$ for which $i_k \neq \bar i_k$ and find the smallest $\bar k>k$ for which $\bar i_{\bar k} = i_k$. If no such $k$ exists, stop.
  \item Due to~\eqref{eq:follow_orbit}, we know that $\PiXstar(\bar i_{\bar k})=\PiXstar(i_k)=\PiXstar([i_{k-1}+1]_{p^\star})=\PiXstar(\bar i_k)$. 
  Hence, the sequence $(\Pistar(\bar i_j))_{j \in \I_{[k, \bar k-1]}}$ is a periodic orbit. 
  Summing up the dissipation inequality~\eqref{eq:strict_diss} from Ass.~\ref{ass:diss} with~\eqref{eq:rot_cost} along this orbit yields 
  \begin{align*}
    0 &\leq \sum_{j=k}^{\bar k-1} \big(\ell(\Pi^\star (\bar i_j)) - \ell^\star \big) =\sum_{j=k}^{\bar k-1} \ell(\Pi^\star (\bar i_j)) - (\bar k - k) \ell^\star
  \end{align*}
  where we used that the terms with the storage function cancel out when summing~\eqref{eq:rot_cost} along an orbit. 
  Hence, the average cost of this $\bar k-k$-periodic orbit is larger than\footnote{It can actually be shown that the cost is always equal to and never larger than $\ell^\star$, but that is not needed here.} or equal to $\ell^\star$.
  \item Remove the points $\I_{[k, \bar k-1]}$ from $\bar i_k$, i.e., redefine $\bar i_j := \bar i_j$ for $j\in \I_{[0,k-1]}$ and $\bar i_j := \bar i_{j-k+\bar k}$ for $j\geq k$. Then, go to 1).
\end{enumerate}
Since we have chosen the length of $\bar i$ initially as $Tp^\star$, we know that we never run out of points in $\bar i$ when applying the above procedure.
However, we may be left with a rather long tail, therefore, if the length of $\bar i$ is longer than or equal to $T+p^\star$, then we remove a multiple of $p^\star$ points at the end of $\bar i$ until we are left with a length between $T$ and $T+p^\star-1$.
The removed part consists solely of full orbits $\Pistar$ and thus has average cost $\ell^\star$.
Finally, $\bar i$ equals the sequence $i$ appended with a tail of maximally $p^\star -1$ more elements.
Since $\bar i$ had initially an average cost of $\ell^\star$ and everything we removed had an average cost larger than or equal to $\ell^ \star$, this means that this remaining sequence has an average cost upper bounded by $\ell^\star$, i.e., 
\begin{align*}
  (T+p^\star -1) \ell^\star \geq \sum_{k=0}^{T-1} \ell(\Pistar(i_k)),
\end{align*}
where we estimated the cost of the tail to be larger than 0 by non-negativity of $\ell$ (compare Rem.~\ref{rem:wlog_ell_nn}).
Together with \eqref{eq:follow_orbit1} this leads to 
\begin{align}\label{eq:follow_orbit2}
  \sum_{k=0}^{T-1} \ell(x_{\hat u}(k,\hat x),\hat u(k)) &\leq (T+ p^\star-1) \ell^\star + T\overline\alpha_\ell(\varepsilon).
\end{align}
for the case $\varepsilon < \varepsilon_1$.

As last step we extend this bound to the case $\varepsilon\geq \varepsilon_1$.
This is straightforward, since the left hand side of the desired inequality is upper bounded by $T \ell_\mathrm{max}$ with $\ell_\mathrm{max} = \sup_{(x,u)\in \X\times \U} \ell(x,u)<\infty$ due to Ass.~\ref{ass:tech}.
Further, using that the stage cost $\ell$ is non-negative, we obtain that the right hand side of the desired inequality is lower bounded by $T\alpha_1(\varepsilon)$.
Hence, when setting $\alpha_1 = \max\left\{\frac{\ell_\mathrm{max}}{\overline\alpha_\ell(\varepsilon_1)}, 1 \right\} \overline\alpha_\ell$, the already established bound \eqref{eq:follow_orbit2} holds for $\alpha_1$ since $\alpha_1(\varepsilon)\geq \overline\alpha_\ell(\varepsilon)$ and further, we obtain $\alpha_1(\varepsilon)\geq \ell_\mathrm{max}$ for $\varepsilon\geq \varepsilon_1$ such that the bound holds as well in this case.  \phantom{q} \hfill $\square$
\end{pf}

\begin{lem}[\VNdiff]\label{lem:V_N_diff}
  Let Ass.~\ref{ass:tech},~\ref{ass:diss}, and~\ref{ass:loc_ctrb} hold. Then, there exists $\delta \in \mathcal L$ and $N_0\in\N$ such that for all $\new{x\in\X_{\alpha,N_0}}$ and all $N\geq N_0$, the following inequality holds
  \begin{align}\label{eq:V_N_diff}
    V_{N+1}^\beta (x)\leq  \frac{N} {N+1} V_{N}^\beta (x) + \ell^\star + \delta(\new{N+1})
  \end{align}
  \new{
  and for $x_1 = x_{\mu_N^\beta}(1, x)$
  \begin{align}\label{eq:rec_feas}
    V_{N}^\beta (x_1)\leq  V_{N}^\beta (x) + \frac{N+1}{N} \left(\ell^\star - \ell(x,\mu_N^\beta(x)) + \delta(N+1)\right).
\end{align}}
\end{lem}
\begin{pf}
We prove this inequality by taking the optimal input $u_{N,x}^\beta$ and using the weak turnpike property $\new{x\in\X_{\alpha,N_0}}$ to construct an almost optimal candidate solution of horizon length $N+1$, with which we can estimate $V_{N+1}^\beta (x)$ \new{and $V_N(x_1)$}.

Let us define the function $\sigma:[N_0,\infty)\to[0,\infty)$ as $\sigma(N):=\alpha^{-1}\big(\sqrt[-4]{N}\big)$.
Since $\sqrt[-4]{N}$ is continuous and monotonically decreasing on $[N_0,\infty)$, it is $\sigma\in\mathcal L$.
\new{As $x \in \X_{\alpha,N_0}$, we can use \eqref{eq:wtp} with} $\varepsilon=\sigma(N)$ \new{to obtain} for all $N\geq N_0$
\begin{align}\label{eq:Q_geq}
  Q_{\sigma(N)}^\beta (N,x) \geq N-\frac{\sqrt{N}}{\alpha(\sigma(N))} = N-\sqrt[4]{N^3}\geq \new{2},
\end{align}
where the last inequality holds if $N_0\geq \new{6}$.
Hence, we are guaranteed that there exists a point $P\in\I_{[\new{1},N-1]}$ in the $\sigma(N)$ neighborhood of $\Pistar$ that satisfies 
\begin{align}\label{eq:P_geq}
  P\geq Q_{\sigma(N)}^\beta (N,x)-1\geq N-\sqrt[4]{N^3}-1,
\end{align}
as there are $Q_{\sigma(N)}^\beta (N,x)$ of such points in $\I_{[0,N-1]}$.
Further, we choose $N_0$ large enough, such that $\sigma(N_0)\leq \kappa$ with $\kappa$ from Ass.~\ref{ass:loc_ctrb}.
Then we know, that $\|x_{u_{N,x}^\beta}(P,x)\|_\PiXstar\leq \kappa$ and that we can choose an input sequence $u'\in\U^{M'}\big(x_{u_{N,x}^\beta}(P,x)\big)$ from Ass.~\ref{ass:loc_ctrb} with $x_{u'}(M',x_{u_{N,x}^\beta}(P,x))=x_{u_{N,x}^\beta}(P,x)$ that keeps the trajectory close to the optimal orbit $\Pistar$ in the sense of \eqref{eq:loc_ctrb} and ends where it started.
Thus, we can fit it as middle piece into the trajectory generated by $u_{N,x}^\beta$ leading to the candidate input sequence over the one step longer horizon $\I_{[0,N]}$
\begin{align*}
  \bar u(k) = \begin{cases}u_{N,x}^\beta (k) & \text{for } k\in \I_{[0,P-1]} \\ 
    u'(k-P) & \text{for } k\in \I_{[P,P+M'-1]} \\ 
    u_{N,x}^\beta (k-M') & \text{for } k\in \I_{[P+M',N]}, \end{cases}
\end{align*}
resulting in the state trajectory
\begin{align*}
  \bar x(k) = \begin{cases}x_{u_{N,x}^\beta} (k,x) & \text{for } k\in \I_{[0,P-1]} \\
    x_{u'}(k-P,x_{u_{N,x}^\beta}(P,x)) & \text{for } k\in \I_{[P,P+M'-1]} \\
    x_{u_{N,x}^\beta}  (k-M',x) & \text{for } k\in \I_{[P+M',N+1]}. \end{cases}
\end{align*}
\new{Note that $\bar x(1)=x_1$ as $P\geq 1$. Hence, this candidate solution gives rise to the upper bounds 
\begin{subequations}\label{eq:VN_bounds}
\begin{align}
  V_{N+1}^\beta (x)&\leq J_{N+1}^\beta (x,\bar u)\\
  V_N^\beta (x_1) &\leq \frac{N+1}{N} J_{N+1}^\beta (x,\bar u) - \frac{N+1}{N}\ell (x,\mu_N^\beta (x)) .
\end{align}
\end{subequations}}
We investigate $J_{N+1}^\beta (x,\bar u)$ in the three parts in which we defined $\bar u$ and $\bar x$.
Let us start with the middle piece $\I_{[P,P+M'-1]}$ as follows
\begin{align*}
  \Sigma_1:&=\sum_{k=P}^{P+M'-1} \beta_{N+1}(k) \ell \left(\bar x(k),\bar u(k)\right)\\
  &\leq \sum_{k=P}^{P+M'-1} \frac{N+1-k}{N+1} \ell_\mathrm{max} \leq M' \frac {N+1-P}{N+1}\ell_\mathrm{max} \\
  & \refeq{\eqref{eq:P_geq}}{\leq}\  M'\ell_\mathrm{max} \left(\frac {2+ \sqrt[4]{N^3}}{{N+1}} \right) \leq M'\ell_\mathrm{max} \left(\frac {2}{N}+\frac{1}{\sqrt[4]{N}} \right) =: \delta_3(N).
\end{align*}
The function $\delta_3$ satisfies $\delta_3\in\mathcal L$.
We continue with $\I_{[0,P-1]}$ and obtain
\begin{align*}
  \Sigma_2:&=\displaystyle\sum_{k=0}^{P-1} \beta_{N+1}(k) \ell \left(\bar x(k),\bar u(k)\right)\\
  &= \sum_{k=0}^{P-1} \frac{N+1-k}{N+1} \ell \left(x_{u_{N,x}^\beta} (k,x),u_{N,x}^\beta (k)\right) \\
  &= \sum_{k=0}^{P-1} \left(\frac{N-k}{N+1}+\frac 1 {N+1} \right) \ell\! \left(x_{u_{N,x}^\beta}\! (k,x),u_{N,x}^\beta (k)\right) \\
  &= \frac{N}{N+1} V_{N}^\beta(x) + \underbrace{\frac 1 {N+1} \sum_{k=0}^{P-1} \ell\! \left(x_{u_{N,x}^\beta}\! (k,x),u_{N,x}^\beta (k)\right)}_{=:\Sigma_4} \\ 
  &\quad - \underbrace{\sum_{k=P}^{N-1} \frac{N-k}{N+1} \ell \left(x_{u_{N,x}^\beta} (k,x),u_{N,x}^\beta (k)\right)}_{=:\Sigma_5}.
\end{align*}
What we gained from this reformulation is that the term $\frac{N}{N+1} V_{N}^\beta(x)$ appears now, which also appears in the bound we want to show \eqref{eq:V_N_diff}.
Before taking care of $\Sigma_4$ and $\Sigma_5$, we take a look at the last piece  $\I_{[P+M',N]}$ 
\begin{align*}
  \Sigma_3:&=\sum_{k=P+M'}^{N} \beta_{N+1}(k) \ell \left(\bar x(k),\bar u(k)\right)\\
  &= \sum_{k=P}^{N-M'} \frac{N+1-M'-k}{N+1} \ell \left(x_{u_{N,x}^\beta} (k,x),u_{N,x}^\beta (k)\right).
\end{align*}
Now that we have named all pieces, let us put them together
\begin{align}\nonumber
  J_{N+1}^\beta(x,\bar u) &= \Sigma_1 + \Sigma_2 + \Sigma_3 \\ \nonumber
  &\leq \frac{N}{N+1} V_{N}^\beta(x) + \delta_3(N)+\Sigma_4-\Sigma_5+\Sigma_3\\
  &\leq \frac{N}{N+1} V_{N}^\beta(x) + \delta_3(N)+\Sigma_4,\label{eq:V_N_diff1}
\end{align}
where we used $\Sigma_3-\Sigma_5\leq 0$, which holds since $\ell$ is non-negative and since the sum $\Sigma_5$ contains more elements and has larger weights than $\Sigma_3$.
This leaves us with $\Sigma_4$, which we can upper bound due to the non-negativity of $\ell$ by
\begin{align*}
  \Sigma_4\leq \frac{1}{N+1} \sum_{k=0}^{N-1} \ell \left(x_{u_{N,x}^\beta}\! (k,x),u_{N,x}^\beta (k)\right).
\end{align*}
In the following we want show that this average (undiscounted) cost of trajectories satisfying the weak turnpike property is approximately $\ell^\star$.
Therefore, we need to use Lemma~\ref{lem:cost_of_traj_at_orbit}, which requires a trajectory fully inside the $\sigma(N)$-neighborhood, however, we have a trajectory that has $N-Q_{\sigma(N)}^\beta (N,x)$ points outside this neighborhood.
Hence, we split the trajectory at these points, such that we end up with $N-Q_{\sigma(N)}^\beta (N,x)+1$ shorter trajectories (possibly of length $0$), which now are fully inside the $\sigma(N)$-neighborhood, as well as the $N-Q_{\sigma(N)}^\beta (N,x)$ outside points.
Each outside point can be conveniently estimated by $\ell_\mathrm{max}$.
Each inside trajectory piece can be estimated with Lemma~\ref{lem:cost_of_traj_at_orbit}.
When we sum up the resulting bounds, the trajectory lengths $T$ in~\eqref{eq:cost_of_traj_at_orbit} of the inside pieces sum up to the total amount of inside points $Q_{\sigma(N)}^\beta (N,x)$, such that we obtain 
\begin{align*}
  \Sigma_4 &\leq \frac{Q_{\sigma(N)}^\beta (N,x)}{N+1} (\ell^\star + \alpha_1(\sigma(N)) ) \\ &\qquad+ \frac{N-Q_{\sigma(N)}^\beta (N,x)+1}{N+1}((p^\star -1) \ell^\star + \ell_\mathrm{max} ).
\end{align*}
Further, we upper bound $\frac{Q_{\sigma(N)}^\beta (N,x)}{N+1}\leq 1$ and 
\begin{align*}
 \frac{N-Q_{\sigma(N)}^\beta (N,x)+1}{N+1} \ \refeq{\eqref{eq:Q_geq}}{\leq}\ \frac{\sqrt[4]{N^3}}{N+1} \leq\frac{\sqrt[4]{N^3}}{N} \leq \frac{1}{\sqrt[4]{N}},
\end{align*}
which leads to
\begin{align*}
  \Sigma_4 &\leq \ell^\star + \underbrace{\alpha_1(\sigma(N)) + \frac{1}{\sqrt[4]{N}}((p^\star -1) \ell^\star + \ell_\mathrm{max} )}_{=:\delta_2(N)}
\end{align*}
where $\delta_2\in\mathcal L$. 
Finally, plugging this bound on $\Sigma_4$ in~\eqref{eq:V_N_diff1} and defining \new{$\delta(N+1):=\delta_3(N)+\delta_2(N)$, $\delta\in\mathcal L$}, we obtain 
\new{
\begin{align*}
  J_{N+1}^\beta (x,\bar u) & \leq \frac{N}{N+1} V_{N}^\beta(x) + \ell^\star + \delta(N+1)
\end{align*}
which yields with~\eqref{eq:VN_bounds} the desired inequalities~\eqref{eq:V_N_diff} and~\eqref{eq:rec_feas}}. \hfill $\square$
\end{pf}

\begin{lem}[\dpp]\label{lem:dpp}\ \\
  Let Ass.~\ref{ass:tech} hold. Then for all $x\in\X$ and all $N\in\N$, the following inequality holds
  \begin{align}\label{eq:dpp}
    V_N^\beta (x) = \ell \Big(x, \mu_{N}^\beta(x)\Big) + \frac {N-1}N V_{N-1}^\beta \Big(f(x,\mu_N^\beta(x))\Big).
  \end{align}
\end{lem}
\begin{pf}
  Consider a general $\bar u\in \U^N(x)$, then we have
  \begin{align}\nonumber
    J_N^\beta (x, \bar u) &= \ell (x,\bar u(0)) + \frac {N-1}N \sum_{k=1}^{N-1} \frac{N-k}{N-1} \ell (x_{\bar u}(k,x),\bar u(k))\\
    &= \ell \Big(x, \bar u(0)\Big) + \frac {N-1}N J_{N-1}^\beta \Big(x_{\bar u} (1,x), \bar u_{[1,N-1]} \Big), \label{eq:dpp_help}
  \end{align}
  where $\bar u_{[1,N-1]} \in\U^{N-1}$ is $\bar u$ without its first element, i.e., $\bar u_{[1,N-1]} (k-1)=\bar u(k)$ for $k\in\I_{[1,N-1]}$.
  Now, we show equality in \eqref{eq:dpp} by showing $\leq$ and $\geq$ starting with $\leq$:
  We take the input sequence defined by $\bar u(0)=\mu_{N}^\beta (x)$ and $\bar u(k)=u_{N-1, x_{\bar u} (1,x)}^\beta (k-1)$ for $k\in \I_{[1,N-1]}$ and use \eqref{eq:dpp_help} as well as $V_N^\beta (x)\leq J_N^\beta (x, \bar u)$ by optimality to prove \eqref{eq:dpp} with $\leq $. 
  For the case $\geq$ we consider a different input sequence: $\bar u = u_{N,x}^\beta$. 
  Using optimality, we obtain $V_{N-1}^\beta \big(x_{\bar u}(1,x)\big) \leq J_{N-1}^\beta (x_{\bar u}(1,x), \bar u_{[1,N-1]})$, which leads with \eqref{eq:dpp_help} to \eqref{eq:dpp} with $\geq$.
  In summary, we have shown $\leq$ and $\geq$, and thus $=$ in \eqref{eq:dpp}. \hfill $\square$
\end{pf}

\begin{lem}[\rotCostZero]\label{lem:rot_cost_zero}
  Let Ass.~\ref{ass:tech},~\ref{ass:diss}, and~\ref{ass:lambda_cont} hold. 
    Then there exist $\underline{\alpha}_{\tilde\ell}, \overline{\alpha}_{\tilde\ell} \in \Kinf$ satisfying for all $(x,u)\in\X\times \U$
  \begin{align}\label{eq:rot_cost_zero}
    \underline\alpha_{\tilde \ell} (\|(x,u)\|_\Pistar) \leq \tilde \ell(x,u) \leq \overline\alpha_{\tilde \ell} (\|(x,u)\|_\Pistar).
  \end{align}
\end{lem}
\begin{pf}
The lower bound is already given by Ass.~\ref{ass:diss}. To prove existence of the upper bound, consider the sum
\begin{align*}
  \tilde \ell ^{p^\star}(\Pistar):=\frac 1 {p^\star} \sum_{k=0}^{p^\star -1} \tilde \ell (\Pistar(k))\refeq{\eqref{eq:rot_cost}}{=}\frac 1 {p^\star} \sum_{k=0}^{p^\star -1} \big(\ell (\Pistar(k))-\ell^\star \big) \refeq{\eqref{eq:ellstar}}{=} 0
\end{align*}
and the fact that $\tilde \ell (\Pistar(k))$ is non-negative for all $k\in \I_{[0,p^\star-1]}$ due to \eqref{eq:strict_diss} from Ass.~\ref{ass:diss}.
If a sum of non-negative summands is zero, we can conclude that all summands are zero, i.e., $\tilde \ell (\Pistar(k))=0$.

Further, continuity of $\ell$ (Ass.~\ref{ass:tech}) and $\lambda$ (Ass.~\ref{ass:lambda_cont}) yields that $\tilde \ell$ is continuous and hence with $\tilde\ell(\Pistar(k))=0$ for all $k\in\I_{[0,p^\star-1]}$ we conclude that there exists $\overline\alpha_{\tilde \ell} \in \Kinf$ satisfying
\begin{align*}
  \tilde \ell(x,u) \leq \overline\alpha_{\tilde \ell} (\|(x,u)\|_\Pistar)
\end{align*}
for all $(x,u)\in \X\times \U$.
\phantom{qed}\hfill $\square$
\end{pf}

\begin{lem}[\rotValUB]\label{lem:rot_val_ub}
  Let Ass.~\ref{ass:tech},~\ref{ass:diss},~\ref{ass:loc_ctrb}, and~\ref{ass:lambda_cont} hold. Then there exists $\overline \alpha_{\tilde V}\in\Kinf$ such that for all $N\in \N$ and all \new{$x\in\X_\mathrm{pi}(C,N_0)$} it holds that
    \begin{align}
      \underline \alpha_{\tilde \ell}(\|(x,\mu_N^\beta(x))\|_\PiXstar) \leq \tilde V_N^\beta (x) \leq \overline \alpha_{\tilde V}(\|x\|_\PiXstar).
  \end{align}
\end{lem}
\begin{pf}
The lower bound on $\tilde V_N^\beta$ is straightforward to see with Lemma~\ref{lem:rot_cost_zero}, since
\begin{align*}
  \tilde V_N^\beta (x) \geq \tilde \ell (x,u_{N,x}^\beta(0)) &\refeq{\eqref{eq:rot_cost_zero}}{\geq} \underline\alpha_{\tilde \ell}(\|(x,u_{N,x}^\beta(0)) \|_\Pistar)\\&\refeq{\eqref{eq:mpc_cl}}{=} \underline\alpha_{\tilde \ell}(\|(x, \mu_N^\beta (x)) \|_\Pistar).
\end{align*}
For the upper bound we recognize due to the positive definiteness of $\tilde \ell$ that $\tilde V_N^\beta(x)\leq \tilde V_{M'+M''}^\beta(x)$ \new{for $N\leq M'+M''$}.
Hence, it is sufficient to prove the upper bound for the case $N>M'+M''$.
For this proof, we make a case distinction.
First, consider $\|x\|_\PiXstar\leq \kappa$ where $\kappa$ is the local controllability neighborhood from Ass.~\ref{ass:loc_ctrb}.
Then, there exists an input sequence $u'\in \U^{M'}(x)$ steering the system in $M'$ steps onto the optimal orbit $\PiXstar$, i.e., satisfying $x_{u'}(M',x)=\PiXstar(l)$ for some $l\in\I_{[0,p^\star -1]}$. 
The candidate input defined by $\bar u(k)=u'(k)$ for $k\in\I_{[0,M'-1]}$ and $\bar u(k)=\PiUstar([k-M'+l]_{p^\star})$ for $k\in \I_{[M',N-1]}$ results in $x_{\bar u}(k)=\PiXstar([k-M'+l]_{p^\star})$ for $k\in \I_{[M',N-1]}$.
Hence, using Lemma~\ref{lem:rot_cost_zero} it follows
\begin{align*}
  \tilde V_N^\beta (x) \leq \tilde J_N^\beta (x,\bar u) &= \sum_{k=0}^{M'-1} \beta_N(k) \tilde\ell(x_{u'}(k,x),u'(k)) \\
  &\refeq{\eqref{eq:loc_ctrb},\,\eqref{eq:rot_cost_zero}}{\leq} \ \ M' \overline\alpha_{\tilde \ell}(\rho  (\|x\|_\PiXstar)).
\end{align*}
For the case $\|x\|_\PiXstar> \kappa$, we use \new{that for $x\in \X_\mathrm{pi}(C,N_0)$ we have $\tilde V_N^\beta (x) \leq \tilde J_N^\beta(x,u_{N,x}^\beta)\leq C$ due to Lemma~\ref{lem:rcf} and~\eqref{eq:Xpi}.} 
To bring both cases together, we define
\begin{align*}
  \overline\alpha_{\tilde V}(s) := \left(\frac{C}{\overline\alpha_{\tilde \ell}(\rho (\kappa))} +M'\right) \overline\alpha_{\tilde \ell}(\rho (s)),
\end{align*}
which satisfies $\overline\alpha_{\tilde V}(\|x\|_\PiXstar)\geq C$ for $\|x\|_\PiXstar\geq \kappa$ as well as $\overline\alpha_{\tilde V}(\|x\|_\PiXstar)\geq M' \overline\alpha_{\tilde \ell}(\rho ( \|x\|_\PiXstar))$.
Hence, we have established the desired bound $\tilde V_N^\beta (x)\leq \overline\alpha_{\tilde V}(\|x\|_\PiXstar)$.
Finally, $\overline\alpha_{\tilde V}\in\Kinf$ is trivially satisfied, since $\left(\frac{C}{\overline\alpha_{\tilde \ell}(\rho (\kappa))} +M'\right)>0$, $\rho \in \Kinf$ and $\overline\alpha_{\tilde \ell}\in\Kinf$, which concludes the proof.
\hfill $\square$
\end{pf}

\begin{lem}[\simcost]\label{lem:sim_cost}\ \\
  Let Ass.~\ref{ass:tech},~\ref{ass:diss},~\ref{ass:loc_ctrb}, and~\ref{ass:lambda_cont} hold and assume that $\Pistar$ is minimal. Then there exists $N_0\in\N$ and $\delta_2\in\mathcal L$ such that for all $x,y\in \new{\X_\mathrm{pi}(C,N_0)}$ and all $N\geq N_0 $ the following inequality holds
    \begin{align}\label{eq:sim_tail_cost}
      &\tilde V_N^\beta (y)-\lambda(y) - \tilde V_N^\beta (x)+\lambda(x) \leq V_N^\beta (y)- V_N^\beta (x) +  \delta_2(N).
  \end{align}
\end{lem}
\begin{pf}
The key insight needed for this proof is that also the rotated optimization problem satisfies the weak turnpike property, which follows \new{from Lemma~\ref{lem:wtp_lb} by checking its assumptions for $\ell$ replaced by $\tilde \ell$:} 
$\tilde \ell$ satisfies the strict dissipativity Ass.~\ref{ass:diss} with $\tilde \lambda =0$ since $\ell$ satisfies this assumption as well. 
\new{Hence, as $\tilde \ell^\star = 0$ (follows from Lemma~\ref{lem:rot_cost_zero} and~\ref{lem:rot_val_ub}) we have by Lemma~\ref{lem:rcf} that the "rotated rotated cost functional" is the same as the rotated cost functional, i.e., $\tilde{\tilde{J}}_N^\beta(x,u)=\tilde{J}_N^\beta(x,u)$.
  Therefore, we have the upper bound $\tilde{\tilde{J}}_N^\beta(x,\tilde u_{N,x}^\beta) = \tilde V_N^\beta(x)\leq \tilde J_N^\beta(x,u_{N,x}^\beta) \leq C$ for all $x\in\X_\mathrm{pi}(C,N_0)$ due to~\eqref{eq:Xpi} and Lemma~\ref{lem:rcf}.
  Therefore, all assumptions of Lemma~\ref{lem:wtp_lb} are satisfied with $\ell$ replaced by $\tilde \ell$ as well and hence, the weak turnpike property from Definition~\ref{defn:wtp} holds also for rotated trajectories starting at $x\in\X_\mathrm{pi}$ with the same $\alpha\in\Kinf$. }

Using the weak turnpike \new{property} \eqref{eq:wtp} for $V_N^\beta (x)$, $V_N^\beta (y)$, $\tilde V_N^\beta (x)$, and $\tilde V_N^\beta (y)$ yields that there are at least $N-\frac{4\sqrt{N}}{\alpha(\varepsilon)}$ time points $k\in \I_{[0,N-1]}$ at which all of these four trajectories are in the $\varepsilon$ neighborhood of $\Pistar$.
We choose $\varepsilon=\sigma_2(N):=\alpha^{-1} (4\sqrt[-4]{N})$ to obtain that there are at least $N-\sqrt[4]{N^3}$ such points.
Let $P\geq N-\sqrt[4]{N^3}-1$ be the largest of these points, which exists, if we choose $N\geq N_0 \geq 4$, i.e., $P\geq 1$.
Further, let $N_0$ be large enough such that $\sigma_2(N_0)\leq \kappa$.
In the following, we will split the proof into two parts, namely we show that $P$ satisfies
\begin{align}\nonumber
  &\tilde V_N^\beta (y)-\lambda(y) - \tilde V_N^\beta (x)+\lambda(x) \\ 
  &\qquad \quad \leq J_P^{\beta_N}(y, u_{N,y}^\beta )-  J_P^{\beta_N}(x, \tilde u_{N,x}^\beta ) +  \delta_4(N)\label{eq:sim_tail_cost1}
\end{align}
and
\begin{align}
  &J_P^{\beta_N}(y, u_{N,y}^\beta )-  J_P^{\beta_N}(x,\tilde u_{N,x}^\beta )\leq  V_N^\beta (y)- V_N^\beta (x)+  \delta_5(N).\label{eq:sim_tail_cost2}
\end{align}
Then, the desired inequality \eqref{eq:sim_tail_cost} immediately follows by defining $\delta_2=\delta_4+\delta_5$.

We start with~\eqref{eq:sim_tail_cost1}. 
Using the local controllability from Ass.~\ref{ass:loc_ctrb}, we can construct a suboptimal candidate solution $\bar u\in\U^{N}(y)$ for $\tilde V_N^\beta (y)$ that follows the optimal solution of $V_N^\beta(y)$ for $P$ steps, then goes in $M'$ steps to the optimal orbit $\Pistar$ to follow it for the remainder of the horizon. 
Thus, we can estimate with Lemma~\ref{lem:rot_cost_zero} ($\tilde \ell ( \Pi^\star (k))=0$) and with~\eqref{eq:loc_ctrb} from the local controllability Assumption~\ref{ass:loc_ctrb} that
\begin{align*}
  \tilde V_N^\beta (y) \leq \tilde J_{N}^{\beta} (y, \bar  u) \leq \tilde J_{P}^{\beta_N} (y, u_{N,y}^\beta) + M'\overline\alpha_{\tilde \ell}(\rho (\sigma_2(N)))
\end{align*}
where we used $\beta_N(k)\leq 1$ to simplify the middle piece.
For notational convenience, we define $\sigma_3(N):=M'\overline\alpha_{\tilde \ell}(\rho (\sigma_2(N)))$, which is $\sigma_3\in\mathcal L$.
Further, since $\tilde \ell$ is non-negative, we can estimate
\begin{align*}
  \tilde J_{P}^{\beta_N} (x, \tilde u_{N,x}^\beta) \leq \tilde V_N^\beta (x).
\end{align*}
Adding these two inequalities, and using Lemma~\ref{lem:rcf} leads to
\begin{align*}
  &\tilde V_N^\beta (y)-\tilde V_N^\beta (x) \\
  &\ \leq \tilde J_{P}^{\beta_N} (y, u_{N,y}^\beta) +  \sigma_3(N)- \tilde J_{P}^{\beta_N} (x, \tilde u_{N,x}^\beta) \\
  &\ \refeq{\eqref{eq:rcf}}{\leq} J_{P}^{\beta_N} (y, u_{N,y}^\beta) +\lambda(y) +  \sigma_3(N)- J_{P}^{\beta_N} (x, \tilde u_{N,x}^\beta)\\
  &\qquad-\lambda(x)+\underbrace{\frac 1 N \sum_{k=1}^{P} \left(\lambda(x_{\tilde u_{N,x}^\beta}(k,x))- \lambda(x_{u_{N,y}^\beta }(k,y))\right)}_{=:\Sigma_6}.
\end{align*}
Let us investigate $\Sigma_6$ in more detail.
Therefore, we use that both $x_{\tilde u_{N,x}^\beta}(k,x)$ and $x_{u_{N,y}^\beta }(k,y)$ satisfy the weak turnpike property and that there are at most $\sqrt[4]{N^3}$ timepoints where one or both trajectories are outside the $\sigma_2(N)$-neighborhood of $\PiXstar$.
When splitting the sum in $\lfloor P/p^\star \rfloor$ parts of length $p^\star$, we know due to the weak turnpike property, that there are at least $\lfloor P/p^\star \rfloor - \sqrt[4]{N^3}$ parts for which both $x_{\tilde u_{N,x}^\beta}(k,x)$ and $x_{u_{N,y}^\beta }(k,y)$ are completely in the $\sigma_2(N)$-neighborhood of $\PiXstar$.
The other $\sqrt[4]{N^3}$ parts and the remaining $[P]_{p^\star}$ points which are too few to have length $p^\star$ must be estimated with the diameter $\overline \lambda=\sup_{x,y\in\X} |\lambda(x)-\lambda(y)|$ on the storage, which is finite as $\X$ is compact (Ass.~\ref{ass:tech}) and $\lambda$ continuous (Ass.~\ref{ass:lambda_cont}).
The parts that are inside the $\sigma_2(N)$-neighborhood of $\Pistar$ can be estimated smarter by exploiting that they follow $\Pistar$, as was proven in \cite[Lemma~15]{Mueller2016}.
For this Lemma, we need that $\Pistar$ is minimal (satisfied by assumption) and we need to choose $N_0$ large enough such that $\sigma_2(N_0)\leq \bar \varepsilon$ with $\bar \varepsilon>0$ from \cite[Lemma~15]{Mueller2016}. 
Hence, for such a part of length $p^\star$ (starting at some $k$) we can estimate with Ass.~\ref{ass:lambda_cont} (continuity of $\lambda$) that
\begin{align*}
  \sum_{j=0}^{p^\star-1} \lambda(x_{\tilde u_{N,x}^\beta}(j+k,x)) &\leq \sum_{j=0}^{p^\star-1} \lambda(\PiXstar(j)) + p^\star \overline \alpha_\lambda (\sigma_2(N))
\end{align*}
and
\begin{align*}
  \sum_{j=0}^{p^\star-1} \lambda(x_{u_{N,y}^\beta}(j+k,y)) &\geq \sum_{j=0}^{p^\star-1} \lambda(\PiXstar(j)) - p^\star \overline \alpha_\lambda (\sigma_2(N)).
\end{align*}
Hence, in the difference of these two sums, the term $\sum_{k=0}^{p^\star-1} \lambda(\PiXstar(k))$ cancels out and $2p^\star \overline \alpha_\lambda (\sigma_2(N))$ remains.
Finally, these arguments lead to
\begin{align*}
  \Sigma_6 &\leq \frac{\bar \lambda(\sqrt[4]{N^3}p^\star + [P]_{p^\star})}{N} + \frac{2p^\star \lfloor P/p^\star \rfloor}N \overline \alpha_\lambda (\sigma_2(N))\\
  &\leq \frac{\bar \lambda p^\star}{\sqrt[4]{N}} + \frac{\bar \lambda (p^\star-1)}{N} + 2 \overline \alpha_\lambda (\sigma_2(N))=:\sigma_4(N)
\end{align*}
and we define $\delta_4(N):=\sigma_3(N)+ \sigma_4(N) $ to conclude \eqref{eq:sim_tail_cost1}.

For the second part of the proof, we must show that \eqref{eq:sim_tail_cost2} is satisfied.
To this end, we construct a suboptimal candidate solution $\bar u\in\U^{N}(x)$ for $V_N^\beta (x)$ that follows the optimal solution $\tilde u_{N,x}^\beta$ of $\tilde V_N^\beta(x)$ for $P$ steps, then follows the input $u'$ from the local controllability of Ass.~\ref{ass:loc_ctrb} that brings the state in $K'\in\I_{[M', M'+p^\star -1]}$ steps\footnote{Since $\tilde x_P:=x_{\tilde u_{N,x}^\beta}(P,x)$ and $y_P:=x_{u_{N,y}^\beta}(P,y)$ are both in the $\sigma_2(N)$ neighborhood of $\PiXstar$, we know that there exist $l_x,l_y\in\I_{[0,p^\star -1]}$ such that $\|\tilde x_P-\PiXstar(l_x)\|\leq \sigma_2(N)$ and $\|y_P-\PiXstar(l_y)\|\leq \sigma_2(N)$. In general, it is not necessarily $l_x=l_y$. Hence, using Ass.~\ref{ass:loc_ctrb} we cannot go directly in $M'$ steps from $\tilde x_P$ to $y_P$ but we first need to follow the orbit for $[l_y-l_x]_{p^\star}$ steps before we can apply the $M'$ local controllability steps.} to $x_{u_{N,y}^\beta}(P,y)$, and then stays on the trajectory of $V_N^\beta (y)$ by applying $u_{N,y}^\beta$. 
Hence,
\begin{align*}
  &V_N^\beta (x) \leq J_N^\beta (x,\bar u)\leq J_P^{\beta_N} (x,\tilde u_{N,x}^\beta) + \\
  & \underbrace{+ \sum_{k=P}^{P+K'-1} \beta_N(k) \ell (x_{u'}\big(k-P,x_{\bar u}(P,y)\big),u'(k-P) )}_{=:\Sigma_7}\\
  &\underbrace{+ \sum_{k=P+K'}^{N-1} \beta_N(k) \ell \big(x_{u_{N,y}^\beta} (k-K',y), u_{N,y}^\beta (k-K')\big) }_{=:\Sigma_8}.
\end{align*}
First, we consider $\Sigma_7$ 
\begin{align*}
  \Sigma_7 &\leq K' \beta_N(P) \ell_\mathrm{max} = \frac{N-P}{N} K'\ell_\mathrm{max} \leq \frac{\sqrt[4]{N^3}+1}{N} K'\ell_\mathrm{max}
\end{align*}
Second, we consider $\Sigma_8$ and use $\beta_N(k)-\beta_N(k-K')=-K'/N$ to obtain
\begin{align*}
  \Sigma_8 &= -\sum_{k=P+K'}^{N-1} \frac{K'}{N} \ell \big(x_{u_{N,y}^\beta} (k-K',y), u_{N,y}^\beta (k-K')\big) + V_N^\beta (y)\\
  &\quad - J_P^{\beta_N} (y,u_{N,y}^\beta )-\!\!\sum_{k=N-K'}^{N-1} \beta_N(k) \ell \big(x_{u_{N,y}^\beta} (k,y), u_{N,y}^\beta (k)\big)\\
  &\leq V_N^\beta (y) - J_P^{\beta_N} (y,u_{N,y}^\beta )
\end{align*}
where we used non-negativity of $\ell$ in the last step.
Bringing the estimates for $\Sigma_7$ and $\Sigma_8$ together and defining $\delta_5(N)=\frac{\sqrt[4]{N^3}+1}{N} K'\ell_\mathrm{max}$, $\delta_5\in\mathcal L$, yields \eqref{eq:sim_tail_cost2} and thus completes the proof.
\phantom{qed}\hfill $\square$
\end{pf}

%% file: root.bbl
\begin{thebibliography}{10}

\bibitem{Amrit2011}
R.~Amrit, J.~B. Rawlings, and D.~Angeli.
\newblock Economic optimization using model predictive control with a terminal
  cost.
\newblock {\em Annual Reviews in Control}, 35(2):178 -- 186, 2011.

\bibitem{Angeli2012}
D.~Angeli, R.~Amrit, and J.~B. Rawlings.
\newblock On average performance and stability of economic model predictive
  control.
\newblock {\em {IEEE} Trans. Automat. Control}, 57(7):1615--1626, 2012.

\bibitem{Berberich2020}
J.~Berberich, J.~Köhler, F.~Allgöwer, and M.~A. Müller.
\newblock Dissipativity properties in constrained optimal control: A
  computational approach.
\newblock {\em Automatica}, 114:108840, 2020.

\bibitem{Brock1972}
W.~A. Brock and L.~J. Mirman.
\newblock Optimal economic growth and uncertainty: The discounted case.
\newblock {\em J. Economic Theory}, 4(3):479--513, 1972.

\bibitem{Broomhead2015}
T.~J. Broomhead, C.~Manzie, R.~C. Shekhar, and P.~Hield.
\newblock Robust periodic economic {MPC} for linear systems.
\newblock {\em Automatica}, 60:30 -- 37, 2015.

\bibitem{Dong2018a}
Z.~Dong and D.~Angeli.
\newblock Analysis of economic model predictive control with terminal penalty
  functions on generalized optimal regimes of operation.
\newblock {\em Int. J. Robust and Nonlinear Control}, 28(16):4790--4815, 2018.

\bibitem{Ellis2014}
M.~Ellis, H.~Durand, and P.~D. Christofides.
\newblock A tutorial review of economic model predictive control methods.
\newblock {\em J. Process Control}, 24(8):1156--1178, 2014.

\bibitem{Faulwasser2015a}
T.~Faulwasser and D.~Bonvin.
\newblock On the design of economic {NMPC} based on an exact turnpike property.
\newblock In {\em Proc. 9th IFAC Symp. Advanced Control of Chemical Processes
  (ADCHEM)}, volume~48, pages 525--530, 2015.

\bibitem{Faulwasser2015b}
T.~Faulwasser and D.~Bonvin.
\newblock On the design of economic {NMPC} based on approximate turnpike
  properties.
\newblock In {\em Proc. 54th {IEEE} Conf. Decision and Control ({CDC})}.
  {IEEE}, 2015.

\bibitem{Faulwasser2018}
T.~Faulwasser, L.~Gr{\"u}ne, and M.~A. M{\"u}ller.
\newblock Economic nonlinear model predictive control.
\newblock {\em Foundations and Trends in Systems and Control}, 5:1--98, 2018.

\bibitem{Gruene2013}
L.~Gr{\"u}ne.
\newblock Economic receding horizon control without terminal constraints.
\newblock {\em Automatica}, 49(3):725 -- 734, 2013.

\bibitem{Gruene2015}
L.~{Gr{\"u}ne} and A.~{Panin}.
\newblock On non-averaged performance of economic {MPC} with terminal
  conditions.
\newblock In {\em Proc. 54th IEEE Conf. Decision and Control (CDC)}, pages
  4332--4337, 2015.

\bibitem{Gruene2014}
L.~Gr{\"u}ne and M.~Stieler.
\newblock Asymptotic stability and transient optimality of economic {MPC}
  without terminal conditions.
\newblock {\em J. Process Control}, 24(8):1187 -- 1196, 2014.

\bibitem{Gruene2022}
L.~Grüne.
\newblock Dissipativity and optimal control: Examining the turnpike phenomenon.
\newblock {\em {IEEE} Control Systems}, 42(2):74--87, 2022.

\bibitem{Gruene2016}
L.~Grüne, C.~M. Kellett, and S.~R. Weller.
\newblock On a discounted notion of strict dissipativity.
\newblock In {\em {IFAC} Symp. Nonlinear Control Systems ({NOLCOS})},
  volume~49, pages 247--252, 2016.

\bibitem{Gruene2021a}
L.~Grüne and L.~Krügel.
\newblock Local turnpike analysis using local dissipativity for discrete time
  discounted optimal control.
\newblock {\em Applied Mathematics {\&} Optimization}, 84(S2):1585--1606, 2021.

\bibitem{Gruene2021}
L.~Grüne, M.~A. Müller, C.~M. Kellett, and S.~R. Weller.
\newblock Strict dissipativity for discrete time discounted optimal control
  problems.
\newblock {\em Mathematical Control {\&} Related Fields}, 11(4):771, 2021.

\bibitem{Gruene2017}
L.~Grüne and J.~Pannek.
\newblock {\em Nonlinear Model Predictive Control: Theory and Algorithms}.
\newblock Springer, Cham, Switzerland, 2017.

\bibitem{Gruene2019}
L.~Grüne and S.~Pirkelmann.
\newblock Economic model predictive control for time-varying system:
  Performance and stability results.
\newblock {\em Optimal Control Applications and Methods}, 41(1):42--64, 2019.

\bibitem{Kloeppelt2021}
C.~Kl{\"o}ppelt, L.~Schwenkel, F.~Allg{\"o}wer, and M.~A. M{\"u}ller.
\newblock Transient performance of tube-based robust economic model predictive
  control.
\newblock In {\em Proc. IFAC Conf. Nonlinear Model Predictive Control (NMPC)},
  pages 28--35, 2021.

\bibitem{Koehler2018}
J.~Köhler, M.~A. Müller, and F.~Allgöwer.
\newblock On periodic dissipativity notions in economic model predictive
  control.
\newblock {\em {IEEE} Control Systems Letters}, 2(3):501--506, 2018.

\bibitem{Martin2019}
T.~Martin, P.~N. K{\"o}hler, and F.~Allg{\"o}wer.
\newblock Dissipativity and economic model predictive control for optimal set
  operation.
\newblock In {\em Proc. American Control Conf. ({ACC})}, 2019.

\bibitem{Mueller2016}
M.~A. Müller and L.~Grüne.
\newblock Economic model predictive control without terminal constraints for
  optimal periodic behavior.
\newblock {\em Automatica}, 70:128--139, 2016.

\bibitem{Mueller2015a}
M.~A. Müller, L.~Grüne, and F.~Allgöwer.
\newblock On the role of dissipativity in economic model predictive control.
\newblock In {\em Proc. 5th {IFAC} Conf. Nonlinear Model Predictive Control
  (NMPC)}, number~23, pages 110--116, 2015.

\bibitem{Schwenkel2020b}
L.~Schwenkel, J.~K{\"o}hler, M.~A. M{\"u}ller, and F.~Allg{\"o}wer.
\newblock Robust economic model predictive control without terminal conditions.
\newblock In {\em Proc. 21st IFAC World Congress}, pages 7097--7104, 2020.

\bibitem{Soloperto2022}
R.~Soloperto, M.~A. Müller, and F.~Allgöwer.
\newblock Guaranteed closed-loop learning in model predictive control.
\newblock {\em IEEE Trans. Automat. Control}, 2022.

\bibitem{Sontag1998a}
Eduardo~D. Sontag.
\newblock {\em Mathematical Control Theory}.
\newblock Springer New York, 1998.

\bibitem{Wabersich2018}
K.~P. Wabersich, F.~A. Bayer, M.~A. M{\"u}ller, and F.~Allg{\"o}wer.
\newblock Economic model predictive control for robust periodic operation with
  guaranteed closed-loop performance.
\newblock In {\em Proc. 16th European Control Conf. ({ECC})}, 2018.

\bibitem{Wuerth2013}
L.~Würth, I.~J. Wolf, and W.~Marquardt.
\newblock On the numerical solution of discounted economic {NMPC} on infinite
  horizons.
\newblock In {\em Proc. 10th IFAC Symp. Dynamics and Control of Process Systems
  (DYCOPS)}, volume~46, pages 209--214, 2013.

\bibitem{Zanon2022}
M.~Zanon and S.~Gros.
\newblock A new dissipativity condition for asymptotic stability of discounted
  economic {MPC}.
\newblock {\em Automatica}, 141:110287, 2022.

\bibitem{Zanon2017}
M.~Zanon, L.~Grüne, and M.~Diehl.
\newblock Periodic optimal control, dissipativity and {MPC}.
\newblock {\em {IEEE} Trans. Automat. Control}, 62(6):2943--2949, 2017.

\end{thebibliography}
